\documentclass[prb, preprint,superscriptaddress,longbibliography]{revtex4-1}

\usepackage{amsmath,amssymb}
\usepackage{graphicx}
\usepackage{epstopdf}
\usepackage{bm}
\usepackage{color}

\newcommand{\as}	{$^{75}$As}
\newcommand{\nlfa}	{Na$_{1-x}$Li$_{x}$FeAs}
\newcommand{\slr} 	{$T_1^{-1}$}
\newcommand{\slrt} 	{$(T_1T)^{-1}$}

\newcommand{\chispin} 	{$\chi_\text{spin}$}

\newcommand{\tsdw} 	{$T_\text{SDW}$}

\newcommand{\tnem} {$T_\text{nem}$}

\newcommand{\bc}[1]{\textbf{\sffamily #1}}

\renewcommand{\figurename}{{\bf Figure}}
\renewcommand{\thefigure}{{\bf \arabic{figure}}}

\begin{document}

\title{Tuning the interplay between nematicity and spin fluctuations in Na$_{1-x}$Li$_x$FeAs superconductors}
\author{S.-H. Baek}
\email[corresponding author: ]{sbaek.fu@gmail.com}
\affiliation{IFW Dresden, Helmholtzstr. 20, 01069 Dresden, Germany}
\author{Dilip Bhoi}
\affiliation{Center for Novel State of Complex Materials Research, Department of Physics and Astronomy, Seoul National
	University, Seoul 151-747, Korea}
\author{Woohyun Nam}
\affiliation{Center for Novel State of Complex Materials Research, Department of Physics and Astronomy, Seoul National
	University, Seoul 151-747, Korea}
\author{Bumsung Lee}
\affiliation{Center for Novel State of Complex Materials Research, Department of Physics and Astronomy, Seoul National
	University, Seoul 151-747, Korea}
\author{D. V. Efremov} 
\affiliation{IFW Dresden, Helmholtzstr. 20, 01069 Dresden, Germany}
\author{B. B\"uchner}
\affiliation{IFW Dresden, Helmholtzstr. 20, 01069 Dresden, Germany}
\affiliation{Department of Physics, Technische Universit\"at Dresden, 01062 Dresden, Germany}
\author{Kee Hoon Kim}
\email[corresponding author: ]{optopia@snu.ac.kr}
\affiliation{Center for Novel State of Complex Materials Research, Department of Physics and Astronomy, Seoul National
	University, Seoul 151-747, Korea}
\affiliation{Institute of Applied Physics, Department of Physics and Astronomy, Seoul National University, Seoul 151-747, Korea}
\date{\today}

\begin{abstract}

Strong interplay of spin and charge/orbital degrees of freedom is the
	fundamental characteristic of the iron-based superconductors (FeSCs),
	which leads to the emergence of a nematic state as a rule in
	the vicinity of the antiferromagnetic state.
Despite intense debate for many years, however,
	whether nematicity is driven by spin or
	orbital fluctuations remains unsettled.
	Here, by use of transport, magnetization, and \as\ nuclear magnetic
	resonance (NMR) measurements, we show a striking
	transformation of the relationship between nematicity and spin
	fluctuations (SFs) in \nlfa;
For $x\leq 0.02$, the nematic transition promotes SFs. In contrast, for $x\geq 0.03$,
the system undergoes a non-magnetic phase
	transition at a temperature $T_0$ into a distinct nematic state that
	suppresses SFs.
	Such a  drastic change of the spin fluctuation spectrum associated with nematicity by small doping is highly unusual, and provides insights into the origin and nature of nematicity in FeSCs.
 
\end{abstract}

\maketitle

\subsection*{Introduction}

Nematicity, i.e., spontaneous breaking of the $C_4$ symmetry of the crystal,
has emerged as a research focus recently in the iron-based superconductors
(FeSCs), because the nematic state can provide a clue to the understanding of high
temperature superconductivity
in these materials\cite{fernandes14,bohmer15a,si16,kuo16,yamakawa16,fernandes17,yi17,bohmer18}.
Currently two major scenarios have been proposed for the origin of nematicity: magnetic and
charge/orbital\cite{chubukov15}. The former assumes that the nematic state
are entirely induced by the interband spin fluctuations (SFs).
The latter scenario treats its origin as charge density
waves or orbital orders.

The magnetic scenario is believed to be realized in the
122-family of FeSCs\cite{fernandes13,lu14,kretzschmar16},
particularly because a scaling relation was found between the spin
fluctuations in nuclear magnetic resonance (NMR)
and the shear modulus in the tetragonal phase of
Ba(Fe$_{1-x}$Co$_x$)$_2$As$_2$ (ref. \onlinecite{fernandes13}).
On the other hand, the most simple compound FeSe
is best described by the orbital scenario since nematic order
occurs without any signature of the spin fluctuation enhancement
\cite{baek15,bohmer15,baek16}.
In FeSCs other than FeSe, however,
the SDW transition temperature \tsdw\ is quite close to
the nematic one \tnem, imposing limitations on investigating the interplay of
nematicity and SFs in detail.
Thus it is much desirable to find a
system in which one can effectively tune SFs and nematicity in wide phase spaces, e.g., temperature and chemical doping.

From this point of view,  NaFeAs, which is
isostructural and isoelectronic
to well investigated LiFeAs, is worth attention.
LiFeAs shows only a superconducting (SC) ground state without a signature of nematicity or magnetism\cite{wang08,tapp08,li10}.
In contrast, NaFeAs is featured by the three successive transitions at low
$T$; a nematic transition at $T_\text{nem}\sim55$ K is followed by a
SDW at $T_\text{SDW}\sim 45$ K
and by a filamentary SC transition at $\sim$ 8 K.
In this respect, the
study of (Na,Li)FeAs may allow a full spectrum of emergent orders coming from a
delicate balance among competing orders by
Li (Na)-substitution into Na (Li) layers in NaFeAs (LiFeAs). However,
the study of such isoelectronic doping has been extremely challenging because
(Na,Li)FeAs becomes easily phase separated due to distinct
chemistry of Na and Li metals.

In this work, we report the successful growth of homogeneous \nlfa\ single
crystals and the
investigation of their electronic phase diagram up to $x=0.1$. We
found that with a systematic increase of $x$, the SDW is suppressed
for $x\geq0.03$, giving way to the SC
state with the full Meissner shielding. Strikingly, also for $x \ge 0.03$,  \as\
spin-lattice relaxation measurements show a sharp anomaly at a well defined
temperature $T_0$, evidencing
a non-magnetic phase transition before entering the bulk SC state. Our
comprehensive data further show that, above a critical doping
$x_\text{c}\sim 0.03$, spin
and nematic fluctuations become strongly entangled, resulting in a charge/orbital ordered state below $T_0$.
This implies that  the nature of a nematic state could vary depending on the underlying electronic structure. Furthermore, our rich phase diagram strongly suggests that  the normal state of FeSCs from which superconductivity emerges is far more complex than previously known.

\subsection*{Results and discussion}

\paragraph*{\bf Crystal structure}
Figure 1a presents the crystal structure and Fig. 1b shows the variation of the $c$-axis lattice parameter of the Na$_{1-x}$Li$_{x}$FeAs single crystals, which decreases systematically with increasing $x$ up to 0.06 and then levels off from 0.08. The $c$ value was extracted from the (00$l$) reflections in the diffraction pattern measured along the $ab$-plane of the single crystals [see Supplementary Fig. 1(a)] which suggests the absence of any other impurity phase. To determine the crystalline phase, we also performed the powder x-ray diffraction of the ground Na$_{0.95}$Li$_{0.05}$FeAs single crystal [see Supplementary Fig. 1(b)] and the pattern could be successfully refined by the tetragonal $P$4/$nmm$ structure as in the parent NaFeAs (ref. \onlinecite{parker09}). 

\paragraph*{\bf Transport and magnetization measurements}
The temperature ($T$) dependence of resistivity ($\rho$) of the selected \nlfa\ single
crystals is displayed in the $T$ range from 3 to 300 K, and near
the SC transition in Fig. 1d and 1e, respectively. Each
resistivity curve was normalized by the value at 300 K ($\rho/\rho_\text{300K}$) and shifted
vertically for clarity (for the original resistivity data, see Supplementary Fig. 2). For the undoped crystal,  $\rho$
decreases smoothly exhibiting a typical metallic
behavior with decreasing $T$, showing up several
anomalous features at low $T$, an upturn at $\sim$54 K (magenta arrow), a first drop (black
arrow) at $\sim$41 K, and a second drop to reach finally zero resistivity state
(blue arrow) at $\sim$7.7 K, which are identified as the nematic (\tnem), the
SDW (\tsdw), and the SC ($T_\text{c}^{\rho}$) transition
temperatures, respectively. \tnem\ and \tsdw\  appear more clearly in
the derivative curves of $d\rho/dT$ as a deviation point (magenta arrow in Fig. 1f)
and a local maximum or minimum (black arrow in Fig. 1f).  At $T_\text{c}^{\rho}\sim$7.7
K, the corresponding magnetic susceptibility ($\chi$) (Fig. 1c) decreases
abruptly, allowing us to assign $T_\text{c}$ from $\chi$,
$T_\text{c}^\chi\sim$7.0 K. These
transport and magnetization data are overall consistent with those found
in previous reports on
NaFeAs (refs. \onlinecite{parker09,parker10a,chen09b,wang12c,wang13,steckel15}).

With the same criteria applied on the parent NaFeAs, we could extract the $T_\text{c}^\rho$
($T_\text{c}^\chi$), \tsdw, \tnem, and Meissner shielding fractions for the whole region
($0\leq x \leq 0.1$). With increasing doping within $0\leq x \leq 0.02$, both
\tnem\ and \tsdw\ are rapidly suppressed.  For $x=0.03$, the $d\rho/dT$ curve
reveals a jump which is assigned to \tnem, but does not show an anomaly associated
with \tsdw. The magnetization data show that the SC volume fraction in the parent compound is very small in agreement with the previous results and increases weakly with doping up to $x\sim0.03$. Upon further doping ($x=0.04$, 0.05, and 0.06), the SC shielding fraction
at $T=2$ K reaches 68, 98, and 90 \%, respectively, and
the highest $T_\text{c}=12.3$ K is obtained for $x= 0.05$, constituting an optimal
doping. After the optimal
doping, inserting more dopants into the system suppress both the SC
shielding fraction and $T_\text{c}$, and eventually superconductivity disappears above $x
= 0.12$. 

\paragraph*{\bf \as\ nuclear magnetic resonance}
We now turn to \as\ NMR measurements on \nlfa.
The \as\ nuclei (nuclear spin $I=3/2$) possess a large quadrupole
moment. For axial symmetry, a
magnetic field $H$ perpendicular to the principal axis (crystallographic $c$-axis
in our case) yields two satellite lines whose separation is given by the quadrupole
frequency $\nu_\text{Q}$.
Figure \ref{spec_ab}a shows the \as\ NMR full spectrum at $H=9.1$ T parallel
to the $ab$ plane as a function of Li doping in the tetragonal/paramagnetic (PM)
phase (at 60 K).
Clearly, $\nu_\text{Q}=9.93$ MHz for undoped crystal does
not change notably with doping, indicating that the local symmetry or the average electric field gradient (EFG) at the \as\ is insensitive to Li dopants up to $x=0.06$.  On the other hand,
the linewidth of the spectra progressively increases with increasing doping,
which is naturally understood by the increase of chemical disorder due to dopants.
It should be noted that the relative linewidth of the satellites with respect
to the central one, which could be considered as a measure of chemical
homogeneity, remains reasonably
small up to $\sim 10$ at $x=0.06$, in support of the high quality of our
samples. Importantly, we do not observe other NMR lines with doping
which would have indicated the presence of an impurity phase such as pure LiFeAs.
Thus, the evolution of the \as\ spectra with doping provides local evidence for
the successful incorporation of Li dopants into the Na layers of NaFeAs.

In the magnetically ordered state of NaFeAs, 
the stripe-like arrangement of Fe moments in the $ab$ plane, by symmetry,
produces hyperfine fields at \as\ pointing along the $c$ axis\cite{kitagawa11,ma11a}.
As a direct consequence of two oppositely aligned antiferromagnetic (AFM)
sublattices, the \as\
central line splits into two AFM lines for $H\parallel c$.  In case of
$H\parallel ab$, the total magnetic field that \as\ experiences slightly
increases due to the vector
sum of the local field along $c$ and the external field along $ab$, and the
magnetic broadening of \as\ line occurs accordingly.
Therefore, long range
AFM order in \nlfa\ can be easily confirmed by NMR via a positive
shift and a broadening of the \as\ line for $H\parallel ab$, and a large
AFM splitting of the line for $H\parallel c$.
Indeed, for the parent ($x=0$) and underdoped ($x=0.02$) samples,
the \as\
line for $H\parallel ab$ broadens and shifts to higher frequency (Fig.
\ref{spec_ab}b and c),
and at the same time the \as\ line for $H\parallel c$ splits into two
well-defined AFM lines (Fig. \ref{spec_c}f), thereby proving locally the SDW order.
Note that the splitting of the \as\ line shown in Fig. \ref{spec_ab}b is due to nemacitiy. 
Indeed, we determined \tnem\ for $x=0$ and 0.02 by measuring the $T$ dependence of \as\ satellite line [see Supplementary Fig. 3].

For $x\ge 0.04$, on the other hand, the $T$ dependence
of the \as\ spectrum is
clearly told apart from those for $x\le 0.02$ samples. First of all, there is
no signature of a static SDW ordering.
The \as\ line preserves its shape without a shift nor a
significant broadening down to low temperatures (see also Supplementary Fig. 4). Secondly, the $T$
evolution of the \as\ spectrum is very similar for the
two field orientations $H\parallel ab$ and $H\parallel c$, which contrasts sharply
with the strongly anisotropic behavior observed for $x\leq 0.02$.  These features
indicate that the system for $x\ge 0.04$ remains paramagnetic.
Remarkably, the intermediate doping, i.e., $x=0.03$, yields a very peculiar
feature which seemingly separates the two doping
regions, $x\leq 0.02$ and $x\geq 0.04$.  That is, the \as\ line
is considerably broadened below $T_0\sim 32$ K for both field orientations
(see Figs. \ref{spec_ab}d and \ref{spec_c}c.). The nearly isotropic NMR line broadening
indicates that an inhomogeneous (short-ranged) magnetism develops. Moreover,
we emphasize that $T_0$
for $x=0.03$ is higher than \tsdw\ for $x=0.02$ and coincides with
\tnem\ (see Fig. \ref{t1t}c). Therefore, we conclude that the inhomogeneous line broadening
observed for $x=0.03$ below $T_0$
is irrelevant to the SDW, but arises from an emerging phase which
may involve a short-ranged magnetism.

For $x\geq 0.03$, alongside the suppression of the SDW, we observe that the resonance
frequency $\nu$ of the \as\ line in the PM phase is abruptly reduced. This
behavior is clearly shown in Fig.
\ref{spec_c}g in terms of the Knight shift,
$\mathcal{K}\equiv (\nu -\nu_0)/\nu_0\times 100$ \% where $\nu_0$
is unshifted Larmor frequency.  Since the second order quadrupole shift vanishes for $H\parallel c$,
$\mathcal{K}_{H\parallel c}$ is equivalent to the local spin susceptibility
$\chi_\text{spin}=\mu_\text{B}^2 N_\text{F}$ where $N_\text{F}$ is the density of states at
the Fermi level.
While a gradual reduction of \chispin\ or $N_\text{F}$ with doping is commonly
observed in FeSCs (refs. \onlinecite{ning10,nakai13,takeda14}),
the abrupt large reduction of \chispin\ induced by a moderate doping
is very unusual [see also Supplementary Fig. 5]. This may suggest a modification of
the Fermi surface geometry near $x\sim 0.03$, owing to, e.g., a
Lifshitz transition\cite{lifshitz60}.

\paragraph*{\bf Low energy spin fluctuations}
Having established that static SDW order is abruptly suppressed at $x\sim0.03$,
we now investigate low energy spin
dynamics, as probed by the spin-lattice relaxation rate divided by temperature
\slrt,  which is proportional to SFs at very low energy.
\slrt\ as a function of $T$ and $x$ are shown in Fig. \ref{t1t}.
For $x=0$ and 0.02,
the diverging behavior of \slrt\ is immediately followed by an exponential
drop with decreasing
$T$. The drastic change of \slrt\ at \tsdw\ precisely reflects two
important characteristics of a SDW transition.  The divergence of \slrt\
at \tsdw\ represents the critical slowing down of SFs toward
long-range magnetic order and the subsequent exponential drop of \slrt\ implies the
depletion of low-energy spin excitations, i.e., the opening of a SDW
gap.
As doping is increased to 0.03, the divergent \slrt\ observed for $x\leq0.02$
is greatly suppressed, being consistent with the disappearance of static SDW order
for $x\geq 0.03$ as discussed above.
Unexpectedly, however, \slrt\ drops rapidly below $T_0>T_\text{c}$ forming a 
peak, indicating a phase transition
at $T_0$. 
With further increasing doping, the \slrt\ peak gradually moves to lower
$T$, but its shape and height remain nearly the same. 

The phase transition at $T_0$ observed for $x\geq0.03$ cannot be of magnetic origin.
Firstly, the $T$ dependence of the \as\ spectrum (Figs. \ref{spec_ab} and
\ref{spec_c}) does not show any signature of static magnetism,
particularly for $x\geq 0.04$.  Secondly, \slrt\ or SFs does not diverge at $T_0$,
implying that the magnetic correlation length remains short at $T_0$.
Thirdly, the sudden reduction of SFs between $x=0.02$ and 0.03
is hardly observed in other FeSCs in which SFs above \tsdw\ is gradually
suppressed with increasing doping or pressure toward the optimal
region\cite{nakai09,ning10,ji13}. 
Moreover, for $x=0.03$, it turns out that  $T_0$ nearly coincides with \tnem, as
shown in Fig. \ref{t1t}c, implying that the $T_0$ transition is closely related to nematicity.
Note that the sharp \slrt\ peak at $T_0$ and its doping dependence are well distinguished from those arising from a glassy spin freezing observed in Co-doped BaFe$_2$As$_2$ (refs. \onlinecite{dioguardi13,dioguardi15}); either significant magnetic broadening or strong doping and field dependence of the \slrt\ peak expected for a glassy magnetic state is not observed. 

Interestingly, we find that the strong anisotropy of \slrt\ above \tsdw\ for $x\leq 0.02$ is maintained above $T_0$ for $x\geq 0.03$ (see Figs. \ref{t1t}a-b). That is, SFs for $H\parallel ab$ is factor of four
stronger than for $H\parallel c$ over the whole doping range investigated. The robust spin fluctuation anisotropy with or without a static SDW order indicates that dynamic SDW fluctuations persist at least up to $x=0.06$.

It may be worthwhile to note that below $T_0$ the \as\ signal amplitude is notably reduced for both $H\parallel ab$ and $H\parallel c$, as shown in Figs. \ref{spec_ab}d-f and \ref{spec_c}c-e (see also Supplementary Fig. 4).
The suppression of signal intensity indicates that the volume fraction of the sample seen by NMR decreases in the charge/orbital ordered phase.  This phenomenon is quite similar to the wipeout effect observed  in the charge stripe phase of cuprate superconductors, in which NMR relaxation rates of the nuclei in spin-rich regions become too fast to be detected\cite{hunt99,imai17,imai18}.
The underlying mechanism of the signal wipeout in \nlfa\ remains unclear and needs further investigation. Interestingly,  the similar wipeout of the NMR signal was also observed in the $^{77}$Se NMR study of FeSe in the nematic phase which does not involve any static magnetism\cite{imai09}.  

\paragraph*{\bf Phase diagram}
Figure \ref{phase} presents the temperature-doping phase diagram determined by
our NMR and transport/magnetization measurements.
When compared to the phase diagram previously known in NaFe$_{1-x}A_x$As
($A$=Co, Cu, or Rh)\cite{wright12,wang13,steckel15,zhou16}, a difference is
the seeming mutual repulsion of the SDW and bulk superconductivity near
$x_\text{c}\sim 0.03$,  similar to that reported in
LaFeAsO$_{1-x}$F$_x$ (ref. \onlinecite{luetkens09}) and NaFe$_{1-x}$Co$_x$As (ref. \onlinecite{ma14}).
However, our data are not sufficiently dense to conclude whether the SDW and SC phases coexist in the narrow doping range near $x=0.03$ or completely repel each other.

The most important feature in Fig. \ref{phase} is the emergence of a non-magnetic phase below $T_0>T_\text{c}$ at optimal doping. Whereas $T_0(x)$ is reasonably connected to \tnem\ for $x\leq 0.02$, we note that the phase below $T_0$ cannot be a simple nematic because the strong suppression of SFs below $T_0$ (or spin-gap behavior) is unlikely due to
nematicity itself.  Although theory predicts that the nematic transition could
trigger a
pseudogap behavior\cite{fernandes12}, such a pseudogap is only viewed as
magnetic precursors whose signature is a sharp increase of the magnetic
correlation length. Therefore, we conclude that the phase below $T_0$
should involve a charge/orbital order which could give rise to a featured gap in the spin fluctuation spectrum.

An even more remarkable observation is the critical change of the relationship
between SFs and nematicity  with doping. For $x\leq 0.02$,
SFs are enhanced just below the nematic transition at \tnem\ and diverge at \tsdw. For
$x\geq 0.03$, however, a strong
enhancement of SFs precedes the phase transition into a charge/orbital nematic state at $T_0$, but it is suppressed once the charge/orbital nematic state develops.
Theoretically, it has been proposed that the strong interplay of spin and
charge/orbital degrees of freedom could result in a charge density wave (CDW) state in proximity to a SDW state
\cite{chubukov08,kang11,chubukov15,classen17}. Following the work\cite{chubukov15},
the competition of these two orders can be described by the effective
Ginzburg-Landau functional:
\begin{eqnarray}
 \Delta F&=& \int_\mathbf{q} \left( \chi_\text{s}^{-1}(\mathbf{q}) ( \mathbf{M}_x^2 + \mathbf{M}_y^2) +  \chi_\text{c}^{-1}(\mathbf{q}) ( \Phi_x^2 + \Phi_y^2)\right) \nonumber \\ &+& \frac{u}{2} \int_\mathbf{q} (\mathbf{M}_x^2 +\mathbf{M}_y^2 +\Phi_x^2 +\Phi_y^2 )^2 - \frac{g}{2} \int_\mathbf{q} (\mathbf{M}_x^2 -\mathbf{M}_y^2 +\Phi_x^2 - \Phi_y^2 )^2,
\end{eqnarray}
where the $\chi_\text{s}(\mathbf{q},\Omega_n) \sim \left[\xi^{-2}_\text{s}+\alpha_\text{s} (q-Q_{x,y})^2\right]^{-1}$
and $\chi_\text{c}(\mathbf{q},\Omega_n)\sim \left[\xi^{-2}_\text{c}+\alpha_\text{c} (q-Q_{x,y})^2\right]^{-1}$ are
dynamical spin and charge susceptibilities, $\mathbf{M}_{x,y}$ and $\Phi_{x,y}$
are SDW and CDW order parameters  with the propagation vectors $Q_x=(\pi,0)$ and
$Q_y=(0,\pi)$, respectively.  The coupling constant $g$ in the leading order arises
due to small non-ellipticity of the electron and hole pockets and is much
smaller than $u$.
The onset of nematic order parameter $\phi$ with
$\mathbf{M}_{x,y}=\Phi_{x,y}=0$  leads to
renormalization of the
magnetic correlation length $\xi^{-2}_s \to \xi^{-2}_s \pm \phi$ (see refs. \onlinecite{fernandes12,chubukov15}). It leads
to strong enhancement of SFs below \tnem. The opposite situation happens
when the nematic transition
coincides with the CDW phase $\Phi_x \neq 0 $. The magnetic correlation length
changes as $\xi_\text{s}^{-2} \to \xi_\text{s}^{-2} + (u\pm g)\Phi_{x}^2 $ leading to
suppression of SFs since $(u\pm g)>0$. 
This may account for the drastic change of the relationship between nematicity and SFs. Our findings establish \nlfa\ as a rich playground for the study of the interplay of spin and charge/orbital degrees of freedom in FeSCs.

\subsection*{Methods}
\paragraph*{\bf Crystal growth and characterization}
High-quality single crystals of \nlfa\ were grown by a self-flux
technique. Due to high reactivity of metallic Na, Li, Fe and As, all
preparation processes were carried out inside an Ar-filled glovebox of which
O$_2$ and H$_2$O contents were less than 1 ppm. Pure elemental Na (99.995\%, Alfa
Aesar), Li (99.9+\%, Sigma Aldrich), As (99.99999+\%, Alfa Aesar) lumps and Fe
(99.998\%, Puratonic) powder in a molar ratio (Na,Li):Fe:As = 3:2:4 were placed
into an alumina crucible, then kept inside a welded Nb container under $\sim$0.8
bar of Ar atmosphere. The welded container was finally sealed in an evacuated
quartz ampoule. The ampoule was heated directly up to 1050 $^\circ$C, stayed at this
temperature for 1 hour, afterward slowly cooled down to 750 $^\circ$C with a
rate of 3 $^\circ$C/h  and then heater was turned off while the ampoule was
still kept in the
furnace. 2D plate-like-shaped single crystals with shiny $ab$-plane surfaces were
mechanically detached from a flux and typical sizes were around $\sim$440.3 mm$^3$.
Since the grown crystals were highly sensitive to air and moisture, each steps
of physical measurement preparations were also done in an Ar-filled glovebox.
The crystalline phase was determined by the powder x-ray diffraction using Cu
K$_\alpha$ radiation at room temperature. To avoid oxidation and compensate the
preferred orientation of single crystals, a sealed quartz capillary was
adopted with grounded as-grown crystals for powder diffraction measurements.
To measure the (00$l$) reflection peak, a piece of single crystal with shiny $ab$-plane surfaces was
	sandwiched between kapton tapes.
Rietveld refinements of the
diffraction pattern were performed using Fullprof program.
In particular, inductively coupled plasma atomic
photoemission spectroscopy (ICP-AES) was carried out on a piece of optimally
doped Na$_{0.95}$Li$_{0.05}$FeAs to check the elemental composition of Na and Li; a
molar ratio of Na and Li was found as Na : Li = 0.944 : 0.056, which is close
to the expected composition.
Electrical resistivity was measured by the conventional four probe technique using a
conductive silver epoxy in PPMS\textsuperscript{TM}
(Quantum Design). Magnetic susceptibility was measured with a vibrating sample magnetometer in PPMS\textsuperscript{TM} and MPMS\textsuperscript{TM}
(Quantum Design).

It should be noted that SC transitions of all the
samples were considerably broad as the estimated transition width $\Delta T$
extracted at temperatures of 10 and 90 \% of resistivity maximum was
7.5 -- 15 K. Moreover, all the resistivity curves exhibited a
temperature window of showing insulating behavior ($d\rho/dT<0$) before the
onset of SC transition similar to NaFe$_{1-x}$Cu$_x$As
system\cite{wang13}. In other doped
Na 111 systems like NaFe$_{1-x}$Co$_x$As (ref. \onlinecite{wang12c}) and
NaFe$_{1-x}$Rh$_x$As (ref. \onlinecite{steckel15}), robust metallic
behavior were observed in broad doping ranges. The former Cu doping
was claimed to have isoelectronic doping while the latter two
involving large electron doping was inevitably accompanied by large
chemical potential shift.  As Li does not bring additional charge
carriers to the systems,
the effect of disorder-induced electron localization on transport
is likely pronounced to result in the insulating-like behaviors as in
the case of NaFe$_{1-x}$Cu$_x$As system.

\paragraph*{\bf Nuclear magnetic resonance} \as\ (nuclear spin $I=3/2$) NMR was carried
out in \nlfa\ single
crystals at an external field of 9.1 T and in
the range of temperature 4.2 -- 100 K.
Due to the extreme sensitivity of the samples to air and moisture, all the samples were
carefully sealed into quartz tubes filled with Ar gas.
The sealed sample was reoriented using a
goniometer for the accurate alignment along the external field. The \as\ NMR
spectra were acquired by a standard spin-echo technique with a typical $\pi/2$
pulse length 2--3 $\mu$s.
The nuclear spin-lattice relaxation rate \slr\ was obtained by fitting the
recovery of the nuclear magnetization $M(t)$ after a saturating pulse to
following fitting function,
$$
1-M(t)/M(\infty)=A[0.9\exp(-6t/T_1)+0.1\exp(-t/T_1)]$$
where $A$ is a fitting parameter. We also measured $^{23}$Na NMR spectra for $x=0.04$.  As shown in Supplementary Figure 6, we confirm that the $^{23}$Na is barely influenced by the $T_0$ phase transition.

\paragraph*{\bf Data Availability} The data that support the findings of this study are available from the corresponding
authors (S.H.B. or K.H.K.).


\begin{thebibliography}{46}%
	\makeatletter
	\providecommand \@ifxundefined [1]{%
		\@ifx{#1\undefined}
	}%
	\providecommand \@ifnum [1]{%
		\ifnum #1\expandafter \@firstoftwo
		\else \expandafter \@secondoftwo
		\fi
	}%
	\providecommand \@ifx [1]{%
		\ifx #1\expandafter \@firstoftwo
		\else \expandafter \@secondoftwo
		\fi
	}%
	\providecommand \natexlab [1]{#1}%
	\providecommand \enquote  [1]{``#1''}%
	\providecommand \bibnamefont  [1]{#1}%
	\providecommand \bibfnamefont [1]{#1}%
	\providecommand \citenamefont [1]{#1}%
	\providecommand \href@noop [0]{\@secondoftwo}%
	\providecommand \href [0]{\begingroup \@sanitize@url \@href}%
	\providecommand \@href[1]{\@@startlink{#1}\@@href}%
	\providecommand \@@href[1]{\endgroup#1\@@endlink}%
	\providecommand \@sanitize@url [0]{\catcode `\\12\catcode `\$12\catcode
		`\&12\catcode `\#12\catcode `\^12\catcode `\_12\catcode `\%12\relax}%
	\providecommand \@@startlink[1]{}%
	\providecommand \@@endlink[0]{}%
	\providecommand \url  [0]{\begingroup\@sanitize@url \@url }%
	\providecommand \@url [1]{\endgroup\@href {#1}{\urlprefix }}%
	\providecommand \urlprefix  [0]{URL }%
	\providecommand \Eprint [0]{\href }%
	\providecommand \doibase [0]{http://dx.doi.org/}%
	\providecommand \selectlanguage [0]{\@gobble}%
	\providecommand \bibinfo  [0]{\@secondoftwo}%
	\providecommand \bibfield  [0]{\@secondoftwo}%
	\providecommand \translation [1]{[#1]}%
	\providecommand \BibitemOpen [0]{}%
	\providecommand \bibitemStop [0]{}%
	\providecommand \bibitemNoStop [0]{.\EOS\space}%
	\providecommand \EOS [0]{\spacefactor3000\relax}%
	\providecommand \BibitemShut  [1]{\csname bibitem#1\endcsname}%
	\let\auto@bib@innerbib\@empty
	\bibitem [{\citenamefont {Fernandes}\ \emph {et~al.}(2014)\citenamefont
		{Fernandes}, \citenamefont {Chubukov},\ and\ \citenamefont
		{Schmalian}}]{fernandes14}%
	\BibitemOpen
	\bibfield  {author} {\bibinfo {author} {\bibfnamefont {R.~M.}\ \bibnamefont
			{Fernandes}}, \bibinfo {author} {\bibfnamefont {A.~V.}\ \bibnamefont
			{Chubukov}}, \ and\ \bibinfo {author} {\bibfnamefont {J.}~\bibnamefont
			{Schmalian}},\ }\bibfield  {title} {\enquote {\bibinfo {title} {What drives
				nematic order in iron-based superconductors?}}\ }\href {\doibase
		10.1038/nphys2877} {\bibfield  {journal} {\bibinfo  {journal} {Nature Phys.}\
		}\textbf {\bibinfo {volume} {10}},\ \bibinfo {pages} {97--104} (\bibinfo
		{year} {2014})}\BibitemShut {NoStop}%
	\bibitem [{\citenamefont {B\"ohmer}\ and\ \citenamefont
		{Meingast}(2015)}]{bohmer15a}%
	\BibitemOpen
	\bibfield  {author} {\bibinfo {author} {\bibfnamefont {A.~E.}\ \bibnamefont
			{B\"ohmer}}\ and\ \bibinfo {author} {\bibfnamefont {C.}~\bibnamefont
			{Meingast}},\ }\bibfield  {title} {\enquote {\bibinfo {title} {{Electronic
					nematic susceptibility of iron-based superconductors}},}\ }\href {\doibase
		10.1016/j.crhy.2015.07.001} {\bibfield  {journal} {\bibinfo  {journal} {C. R.
				Physique}\ }\textbf {\bibinfo {volume} {17}},\ \bibinfo {pages} {90}
		(\bibinfo {year} {2015})}\BibitemShut {NoStop}%
	\bibitem [{\citenamefont {Si}\ \emph {et~al.}(2016)\citenamefont {Si},
		\citenamefont {Yu},\ and\ \citenamefont {Abrahams}}]{si16}%
	\BibitemOpen
	\bibfield  {author} {\bibinfo {author} {\bibfnamefont {Q.}~\bibnamefont
			{Si}}, \bibinfo {author} {\bibfnamefont {R.}~\bibnamefont {Yu}}, \ and\
		\bibinfo {author} {\bibfnamefont {E.}~\bibnamefont {Abrahams}},\ }\bibfield
	{title} {\enquote {\bibinfo {title} {{High-temperature superconductivity in
					iron pnictides and chalcogenides}},}\ }\href {\doibase
		10.1038/natrevmats.2016.17} {\bibfield  {journal} {\bibinfo  {journal} {Nat.
				Rev. Mater.}\ }\textbf {\bibinfo {volume} {1}},\ \bibinfo {pages} {16017}
		(\bibinfo {year} {2016})}\BibitemShut {NoStop}%
	\bibitem [{\citenamefont {Kuo}\ \emph {et~al.}(2016)\citenamefont {Kuo},
		\citenamefont {Chu}, \citenamefont {Palmstrom}, \citenamefont {Kivelson},\
		and\ \citenamefont {Fisher}}]{kuo16}%
	\BibitemOpen
	\bibfield  {author} {\bibinfo {author} {\bibfnamefont {H.-H.}\ \bibnamefont
			{Kuo}}, \bibinfo {author} {\bibfnamefont {J.-H.}\ \bibnamefont {Chu}},
		\bibinfo {author} {\bibfnamefont {J.~C.}\ \bibnamefont {Palmstrom}}, \bibinfo
		{author} {\bibfnamefont {S.~A.}\ \bibnamefont {Kivelson}}, \ and\ \bibinfo
		{author} {\bibfnamefont {I.~R.}\ \bibnamefont {Fisher}},\ }\bibfield  {title}
	{\enquote {\bibinfo {title} {{Ubiquitous signatures of nematic quantum
					criticality in optimally doped Fe-based superconductors}},}\ }\href {\doibase
		10.1126/science.aab0103} {\bibfield  {journal} {\bibinfo  {journal}
			{Science}\ }\textbf {\bibinfo {volume} {352}},\ \bibinfo {pages} {958--962}
		(\bibinfo {year} {2016})}\BibitemShut {NoStop}%
	\bibitem [{\citenamefont {Yamakawa}\ \emph {et~al.}(2016)\citenamefont
		{Yamakawa}, \citenamefont {Onari},\ and\ \citenamefont
		{Kontani}}]{yamakawa16}%
	\BibitemOpen
	\bibfield  {author} {\bibinfo {author} {\bibfnamefont {Y.}~\bibnamefont
			{Yamakawa}}, \bibinfo {author} {\bibfnamefont {S.}~\bibnamefont {Onari}}, \
		and\ \bibinfo {author} {\bibfnamefont {H.}~\bibnamefont {Kontani}},\
	}\bibfield  {title} {\enquote {\bibinfo {title} {{Nematicity and Magnetism in
					FeSe and Other Families of Fe-Based Superconductors}},}\ }\href {\doibase
		10.1103/PhysRevX.6.021032} {\bibfield  {journal} {\bibinfo  {journal} {Phys.
				Rev. X}\ }\textbf {\bibinfo {volume} {6}},\ \bibinfo {pages} {021032}
		(\bibinfo {year} {2016})}\BibitemShut {NoStop}%
	\bibitem [{\citenamefont {Fernandes}\ and\ \citenamefont
		{Chubukov}(2017)}]{fernandes17}%
	\BibitemOpen
	\bibfield  {author} {\bibinfo {author} {\bibfnamefont {R.~M.}\ \bibnamefont
			{Fernandes}}\ and\ \bibinfo {author} {\bibfnamefont {A.~V.}\ \bibnamefont
			{Chubukov}},\ }\bibfield  {title} {\enquote {\bibinfo {title} {Low-energy
				microscopic models for iron-based superconductors: a review},}\ }\href
	{\doibase 10.1088/1361-6633/80/1/014503} {\bibfield  {journal} {\bibinfo
			{journal} {Rep. Prog. Phys.}\ }\textbf {\bibinfo {volume} {80}},\ \bibinfo
		{pages} {014503} (\bibinfo {year} {2017})}\BibitemShut {NoStop}%
	\bibitem [{\citenamefont {Yi}\ \emph {et~al.}(2017)\citenamefont {Yi},
		\citenamefont {Zhang}, \citenamefont {Shen},\ and\ \citenamefont
		{Lu}}]{yi17}%
	\BibitemOpen
	\bibfield  {author} {\bibinfo {author} {\bibfnamefont {Ming}\ \bibnamefont
			{Yi}}, \bibinfo {author} {\bibfnamefont {Yan}\ \bibnamefont {Zhang}},
		\bibinfo {author} {\bibfnamefont {Zhi-Xun}\ \bibnamefont {Shen}}, \ and\
		\bibinfo {author} {\bibfnamefont {Donghui}\ \bibnamefont {Lu}},\ }\bibfield
	{title} {\enquote {\bibinfo {title} {Role of the orbital degree of freedom in
				iron-based superconductors},}\ }\href {\doibase 10.1038/s41535-017-0059-y}
	{\bibfield  {journal} {\bibinfo  {journal} {Quantum Materials}\ }\textbf
		{\bibinfo {volume} {2}},\ \bibinfo {pages} {57} (\bibinfo {year}
		{2017})}\BibitemShut {NoStop}%
	\bibitem [{\citenamefont {B\"ohmer}\ and\ \citenamefont
		{Kreisel}(2018)}]{bohmer18}%
	\BibitemOpen
	\bibfield  {author} {\bibinfo {author} {\bibfnamefont {A.~E.}\ \bibnamefont
			{B\"ohmer}}\ and\ \bibinfo {author} {\bibfnamefont {A.}~\bibnamefont
			{Kreisel}},\ }\bibfield  {title} {\enquote {\bibinfo {title} {{Nematicity,
					magnetism and superconductivity in FeSe}},}\ }\href@noop {} {\bibfield
		{journal} {\bibinfo  {journal} {J. Phys.: Condens. Matter}\ }\textbf
		{\bibinfo {volume} {30}},\ \bibinfo {pages} {023001} (\bibinfo {year}
		{2018})}\BibitemShut {NoStop}%
	\bibitem [{\citenamefont {Chubukov}\ \emph {et~al.}(2015)\citenamefont
		{Chubukov}, \citenamefont {Fernandes},\ and\ \citenamefont
		{Schmalian}}]{chubukov15}%
	\BibitemOpen
	\bibfield  {author} {\bibinfo {author} {\bibfnamefont {A.~V.}\ \bibnamefont
			{Chubukov}}, \bibinfo {author} {\bibfnamefont {R.~M.}\ \bibnamefont
			{Fernandes}}, \ and\ \bibinfo {author} {\bibfnamefont {J.}~\bibnamefont
			{Schmalian}},\ }\bibfield  {title} {\enquote {\bibinfo {title} {{Origin of
					nematic order in FeSe}},}\ }\href {\doibase 10.1103/PhysRevB.91.201105}
	{\bibfield  {journal} {\bibinfo  {journal} {Phys. Rev. B}\ }\textbf {\bibinfo
			{volume} {91}},\ \bibinfo {pages} {201105} (\bibinfo {year}
		{2015})}\BibitemShut {NoStop}%
	\bibitem [{\citenamefont {Fernandes}\ \emph {et~al.}(2013)\citenamefont
		{Fernandes}, \citenamefont {B\"ohmer}, \citenamefont {Meingast},\ and\
		\citenamefont {Schmalian}}]{fernandes13}%
	\BibitemOpen
	\bibfield  {author} {\bibinfo {author} {\bibfnamefont {R.~M.}\ \bibnamefont
			{Fernandes}}, \bibinfo {author} {\bibfnamefont {A.~E.}\ \bibnamefont
			{B\"ohmer}}, \bibinfo {author} {\bibfnamefont {C.}~\bibnamefont {Meingast}},
		\ and\ \bibinfo {author} {\bibfnamefont {J.}~\bibnamefont {Schmalian}},\
	}\bibfield  {title} {\enquote {\bibinfo {title} {{Scaling between Magnetic
					and Lattice Fluctuations in Iron Pnictide Superconductors}},}\ }\href
	{\doibase 10.1103/PhysRevLett.111.137001} {\bibfield  {journal} {\bibinfo
			{journal} {Phys. Rev. Lett.}\ }\textbf {\bibinfo {volume} {111}},\ \bibinfo
		{pages} {137001} (\bibinfo {year} {2013})}\BibitemShut {NoStop}%
	\bibitem [{\citenamefont {Lu}\ \emph {et~al.}(2014)\citenamefont {Lu},
		\citenamefont {Park}, \citenamefont {Zhang}, \citenamefont {Luo},
		\citenamefont {Nevidomskyy}, \citenamefont {Si},\ and\ \citenamefont
		{Dai}}]{lu14}%
	\BibitemOpen
	\bibfield  {author} {\bibinfo {author} {\bibfnamefont {X.}~\bibnamefont
			{Lu}}, \bibinfo {author} {\bibfnamefont {J.~T.}\ \bibnamefont {Park}},
		\bibinfo {author} {\bibfnamefont {R.}~\bibnamefont {Zhang}}, \bibinfo
		{author} {\bibfnamefont {H.}~\bibnamefont {Luo}}, \bibinfo {author}
		{\bibfnamefont {A.~H.}\ \bibnamefont {Nevidomskyy}}, \bibinfo {author}
		{\bibfnamefont {Q.}~\bibnamefont {Si}}, \ and\ \bibinfo {author}
		{\bibfnamefont {P.}~\bibnamefont {Dai}},\ }\bibfield  {title} {\enquote
		{\bibinfo {title} {{Nematic spin correlations in the tetragonal state of
					uniaxial-strained BaFe$_{2-x}$Ni$_x$As$_2$}},}\ }\href {\doibase
		10.1126/science.1251853} {\bibfield  {journal} {\bibinfo  {journal}
			{Science}\ }\textbf {\bibinfo {volume} {345}},\ \bibinfo {pages} {657--660}
		(\bibinfo {year} {2014})}\BibitemShut {NoStop}%
	\bibitem [{\citenamefont {Kretzschmar}\ \emph {et~al.}(2016)\citenamefont
		{Kretzschmar}, \citenamefont {B\"ohm}, \citenamefont {Karahasanović},
		\citenamefont {Muschler}, \citenamefont {Baum}, \citenamefont {Jost},
		\citenamefont {Schmalian}, \citenamefont {Caprara}, \citenamefont {Grilli},
		\citenamefont {Di~Castro}, \citenamefont {Analytis}, \citenamefont {Chu},
		\citenamefont {Fisher},\ and\ \citenamefont {Hackl}}]{kretzschmar16}%
	\BibitemOpen
	\bibfield  {author} {\bibinfo {author} {\bibfnamefont {F.}~\bibnamefont
			{Kretzschmar}}, \bibinfo {author} {\bibfnamefont {T.}~\bibnamefont {B\"ohm}},
		\bibinfo {author} {\bibfnamefont {U.}~\bibnamefont {Karahasanović}},
		\bibinfo {author} {\bibfnamefont {B.}~\bibnamefont {Muschler}}, \bibinfo
		{author} {\bibfnamefont {A.}~\bibnamefont {Baum}}, \bibinfo {author}
		{\bibfnamefont {D.}~\bibnamefont {Jost}}, \bibinfo {author} {\bibfnamefont
			{J.}~\bibnamefont {Schmalian}}, \bibinfo {author} {\bibfnamefont
			{S.}~\bibnamefont {Caprara}}, \bibinfo {author} {\bibfnamefont
			{M.}~\bibnamefont {Grilli}}, \bibinfo {author} {\bibfnamefont
			{C.}~\bibnamefont {Di~Castro}}, \bibinfo {author} {\bibfnamefont {J.~G.}\
			\bibnamefont {Analytis}}, \bibinfo {author} {\bibfnamefont {J.-H.}\
			\bibnamefont {Chu}}, \bibinfo {author} {\bibfnamefont {I.~R.}\ \bibnamefont
			{Fisher}}, \ and\ \bibinfo {author} {\bibfnamefont {R.}~\bibnamefont
			{Hackl}},\ }\bibfield  {title} {\enquote {\bibinfo {title} {{Critical spin
					fluctuations and the origin of nematic order in
					Ba(Fe$_{1-x}$Co$_x$)$_2$As$_2$}},}\ }\href {\doibase 10.1038/nphys3634}
	{\bibfield  {journal} {\bibinfo  {journal} {Nature Phys.}\ }\textbf {\bibinfo
			{volume} {12}},\ \bibinfo {pages} {560} (\bibinfo {year} {2016})}\BibitemShut
	{NoStop}%
	\bibitem [{\citenamefont {Baek}\ \emph {et~al.}(2015)\citenamefont {Baek},
		\citenamefont {Efremov}, \citenamefont {Ok}, \citenamefont {Kim},
		\citenamefont {van~den Brink},\ and\ \citenamefont {B\"uchner}}]{baek15}%
	\BibitemOpen
	\bibfield  {author} {\bibinfo {author} {\bibfnamefont {S.-H.}\ \bibnamefont
			{Baek}}, \bibinfo {author} {\bibfnamefont {D.~V.}\ \bibnamefont {Efremov}},
		\bibinfo {author} {\bibfnamefont {J.~M.}\ \bibnamefont {Ok}}, \bibinfo
		{author} {\bibfnamefont {J.~S.}\ \bibnamefont {Kim}}, \bibinfo {author}
		{\bibfnamefont {J.}~\bibnamefont {van~den Brink}}, \ and\ \bibinfo {author}
		{\bibfnamefont {B.}~\bibnamefont {B\"uchner}},\ }\bibfield  {title} {\enquote
		{\bibinfo {title} {{Orbital-driven nematicity in FeSe}},}\ }\href {\doibase
		10.1038/nmat4138} {\bibfield  {journal} {\bibinfo  {journal} {Nature Mater.}\
		}\textbf {\bibinfo {volume} {14}},\ \bibinfo {pages} {210--214} (\bibinfo
		{year} {2015})}\BibitemShut {NoStop}%
	\bibitem [{\citenamefont {B\"ohmer}\ \emph {et~al.}(2015)\citenamefont
		{B\"ohmer}, \citenamefont {Arai}, \citenamefont {Hardy}, \citenamefont
		{Hattori}, \citenamefont {Iye}, \citenamefont {Wolf}, \citenamefont
		{L\"ohneysen}, \citenamefont {Ishida},\ and\ \citenamefont
		{Meingast}}]{bohmer15}%
	\BibitemOpen
	\bibfield  {author} {\bibinfo {author} {\bibfnamefont {A.~E.}\ \bibnamefont
			{B\"ohmer}}, \bibinfo {author} {\bibfnamefont {T.}~\bibnamefont {Arai}},
		\bibinfo {author} {\bibfnamefont {F.}~\bibnamefont {Hardy}}, \bibinfo
		{author} {\bibfnamefont {T.}~\bibnamefont {Hattori}}, \bibinfo {author}
		{\bibfnamefont {T.}~\bibnamefont {Iye}}, \bibinfo {author} {\bibfnamefont
			{T.}~\bibnamefont {Wolf}}, \bibinfo {author} {\bibfnamefont {H.~v.}\
			\bibnamefont {L\"ohneysen}}, \bibinfo {author} {\bibfnamefont
			{K.}~\bibnamefont {Ishida}}, \ and\ \bibinfo {author} {\bibfnamefont
			{C.}~\bibnamefont {Meingast}},\ }\bibfield  {title} {\enquote {\bibinfo
			{title} {{Origin of the Tetragonal-to-Orthorhombic Phase Transition in FeSe:
					A Combined Thermodynamic and NMR Study of Nematicity}},}\ }\href {\doibase
		10.1103/PhysRevLett.114.027001} {\bibfield  {journal} {\bibinfo  {journal}
			{Phys. Rev. Lett.}\ }\textbf {\bibinfo {volume} {114}},\ \bibinfo {pages}
		{027001} (\bibinfo {year} {2015})}\BibitemShut {NoStop}%
	\bibitem [{\citenamefont {Baek}\ \emph {et~al.}(2016)\citenamefont {Baek},
		\citenamefont {Efremov}, \citenamefont {Ok}, \citenamefont {Kim},
		\citenamefont {van~den Brink},\ and\ \citenamefont {B\"uchner}}]{baek16}%
	\BibitemOpen
	\bibfield  {author} {\bibinfo {author} {\bibfnamefont {S.-H.}\ \bibnamefont
			{Baek}}, \bibinfo {author} {\bibfnamefont {D.~V.}\ \bibnamefont {Efremov}},
		\bibinfo {author} {\bibfnamefont {J.~M.}\ \bibnamefont {Ok}}, \bibinfo
		{author} {\bibfnamefont {J.~S.}\ \bibnamefont {Kim}}, \bibinfo {author}
		{\bibfnamefont {J.}~\bibnamefont {van~den Brink}}, \ and\ \bibinfo {author}
		{\bibfnamefont {B.}~\bibnamefont {B\"uchner}},\ }\bibfield  {title} {\enquote
		{\bibinfo {title} {{Nematicity and in-plane anisotropy of superconductivity
					in $\ensuremath{\beta}-\mathrm{FeSe}$ detected by $^{77}\mathrm{Se}$ nuclear
					magnetic resonance}},}\ }\href {\doibase 10.1103/PhysRevB.93.180502}
	{\bibfield  {journal} {\bibinfo  {journal} {Phys. Rev. B}\ }\textbf {\bibinfo
			{volume} {93}},\ \bibinfo {pages} {180502} (\bibinfo {year}
		{2016})}\BibitemShut {NoStop}%
	\bibitem [{\citenamefont {Wang}\ \emph {et~al.}(2008)\citenamefont {Wang},
		\citenamefont {Liu}, \citenamefont {Lv}, \citenamefont {Gao}, \citenamefont
		{Yang}, \citenamefont {Yu}, \citenamefont {Li},\ and\ \citenamefont
		{Jin}}]{wang08}%
	\BibitemOpen
	\bibfield  {author} {\bibinfo {author} {\bibfnamefont {X.C.}\ \bibnamefont
			{Wang}}, \bibinfo {author} {\bibfnamefont {Q.Q.}\ \bibnamefont {Liu}},
		\bibinfo {author} {\bibfnamefont {Y.X.}\ \bibnamefont {Lv}}, \bibinfo
		{author} {\bibfnamefont {W.B.}\ \bibnamefont {Gao}}, \bibinfo {author}
		{\bibfnamefont {L.X.}\ \bibnamefont {Yang}}, \bibinfo {author} {\bibfnamefont
			{R.C.}\ \bibnamefont {Yu}}, \bibinfo {author} {\bibfnamefont {F.Y.}\
			\bibnamefont {Li}}, \ and\ \bibinfo {author} {\bibfnamefont {C.Q.}\
			\bibnamefont {Jin}},\ }\bibfield  {title} {\enquote {\bibinfo {title} {{The
					superconductivity at 18 K in LiFeAs system}},}\ }\href {\doibase
		10.1016/j.ssc.2008.09.057} {\bibfield  {journal} {\bibinfo  {journal} {Solid
				State Commun.}\ }\textbf {\bibinfo {volume} {148}},\ \bibinfo {pages} {538 --
			540} (\bibinfo {year} {2008})}\BibitemShut {NoStop}%
	\bibitem [{\citenamefont {Tapp}\ \emph {et~al.}(2008)\citenamefont {Tapp},
		\citenamefont {Tang}, \citenamefont {Lv}, \citenamefont {Sasmal},
		\citenamefont {Lorenz}, \citenamefont {Chu},\ and\ \citenamefont
		{Guloy}}]{tapp08}%
	\BibitemOpen
	\bibfield  {author} {\bibinfo {author} {\bibfnamefont {J.~H.}\ \bibnamefont
			{Tapp}}, \bibinfo {author} {\bibfnamefont {Z.}~\bibnamefont {Tang}}, \bibinfo
		{author} {\bibfnamefont {B.}~\bibnamefont {Lv}}, \bibinfo {author}
		{\bibfnamefont {K.}~\bibnamefont {Sasmal}}, \bibinfo {author} {\bibfnamefont
			{B.}~\bibnamefont {Lorenz}}, \bibinfo {author} {\bibfnamefont {P.~C.~W.}\
			\bibnamefont {Chu}}, \ and\ \bibinfo {author} {\bibfnamefont {A.~M.}\
			\bibnamefont {Guloy}},\ }\bibfield  {title} {\enquote {\bibinfo {title}
			{{LiFeAs: An intrinsic FeAs-based superconductor with $T_c=18$ K}},}\ }\href
	{\doibase 10.1103/PhysRevB.78.060505} {\bibfield  {journal} {\bibinfo
			{journal} {Phys. Rev. B}\ }\textbf {\bibinfo {volume} {78}},\ \bibinfo
		{pages} {060505} (\bibinfo {year} {2008})}\BibitemShut {NoStop}%
	\bibitem [{\citenamefont {Li}\ \emph {et~al.}(2010)\citenamefont {Li},
		\citenamefont {Ooe}, \citenamefont {Wang}, \citenamefont {Liu}, \citenamefont
		{Jin}, \citenamefont {M.Ichioka},\ and\ \citenamefont {q.~Zheng}}]{li10}%
	\BibitemOpen
	\bibfield  {author} {\bibinfo {author} {\bibfnamefont {Z.}~\bibnamefont
			{Li}}, \bibinfo {author} {\bibfnamefont {Y.}~\bibnamefont {Ooe}}, \bibinfo
		{author} {\bibfnamefont {X.-C.}\ \bibnamefont {Wang}}, \bibinfo {author}
		{\bibfnamefont {Q.-Q.}\ \bibnamefont {Liu}}, \bibinfo {author} {\bibfnamefont
			{C.-Q.}\ \bibnamefont {Jin}}, \bibinfo {author} {\bibnamefont {M.Ichioka}}, \
		and\ \bibinfo {author} {\bibfnamefont {G.}~\bibnamefont {q.~Zheng}},\
	}\bibfield  {title} {\enquote {\bibinfo {title} {{$^{75}$As NQR and NMR
					studies of superconductivity and electron correlations in iron arsenide
					LiFeAs}},}\ }\href {\doibase 10.1143/JPSJ.79.083702} {\bibfield  {journal}
		{\bibinfo  {journal} {J. Phys. Soc. Jpn.}\ }\textbf {\bibinfo {volume}
			{79}},\ \bibinfo {pages} {083702} (\bibinfo {year} {2010})}\BibitemShut
	{NoStop}%
	\bibitem [{\citenamefont {Parker}\ \emph {et~al.}(2009)\citenamefont {Parker},
		\citenamefont {Pitcher}, \citenamefont {Baker}, \citenamefont {Franke},
		\citenamefont {Lancaster}, \citenamefont {Blundell},\ and\ \citenamefont
		{Clarke}}]{parker09}%
	\BibitemOpen
	\bibfield  {author} {\bibinfo {author} {\bibfnamefont {D.~R.}\ \bibnamefont
			{Parker}}, \bibinfo {author} {\bibfnamefont {M.~J.}\ \bibnamefont {Pitcher}},
		\bibinfo {author} {\bibfnamefont {P.~J.}\ \bibnamefont {Baker}}, \bibinfo
		{author} {\bibfnamefont {I.}~\bibnamefont {Franke}}, \bibinfo {author}
		{\bibfnamefont {T.}~\bibnamefont {Lancaster}}, \bibinfo {author}
		{\bibfnamefont {S.~J.}\ \bibnamefont {Blundell}}, \ and\ \bibinfo {author}
		{\bibfnamefont {S.~J.}\ \bibnamefont {Clarke}},\ }\bibfield  {title}
	{\enquote {\bibinfo {title} {{Structure, antiferromagnetism and
					superconductivity of the layered iron arsenide NaFeAs}},}\ }\href {\doibase
		10.1039/B818911K} {\bibfield  {journal} {\bibinfo  {journal} {Chem. Commun.}\
			,\ \bibinfo {pages} {2189--2191}} (\bibinfo {year} {2009})}\BibitemShut
	{NoStop}%
	\bibitem [{\citenamefont {Parker}\ \emph {et~al.}(2010)\citenamefont {Parker},
		\citenamefont {Smith}, \citenamefont {Lancaster}, \citenamefont {Steele},
		\citenamefont {Franke}, \citenamefont {Baker}, \citenamefont {Pratt},
		\citenamefont {Pitcher}, \citenamefont {Blundell},\ and\ \citenamefont
		{Clarke}}]{parker10a}%
	\BibitemOpen
	\bibfield  {author} {\bibinfo {author} {\bibfnamefont {D.~R.}\ \bibnamefont
			{Parker}}, \bibinfo {author} {\bibfnamefont {M.~J.~P.}\ \bibnamefont
			{Smith}}, \bibinfo {author} {\bibfnamefont {T.}~\bibnamefont {Lancaster}},
		\bibinfo {author} {\bibfnamefont {A.~J.}\ \bibnamefont {Steele}}, \bibinfo
		{author} {\bibfnamefont {I.}~\bibnamefont {Franke}}, \bibinfo {author}
		{\bibfnamefont {P.~J.}\ \bibnamefont {Baker}}, \bibinfo {author}
		{\bibfnamefont {F.~L.}\ \bibnamefont {Pratt}}, \bibinfo {author}
		{\bibfnamefont {M.~J.}\ \bibnamefont {Pitcher}}, \bibinfo {author}
		{\bibfnamefont {S.~J.}\ \bibnamefont {Blundell}}, \ and\ \bibinfo {author}
		{\bibfnamefont {S.~J.}\ \bibnamefont {Clarke}},\ }\bibfield  {title}
	{\enquote {\bibinfo {title} {{Control of the Competition between a Magnetic
					Phase and a Superconducting Phase in Cobalt-Doped and Nickel-Doped NaFeAs
					Using Electron Count}},}\ }\href {\doibase 10.1103/PhysRevLett.104.057007}
	{\bibfield  {journal} {\bibinfo  {journal} {Phys. Rev. Lett.}\ }\textbf
		{\bibinfo {volume} {104}},\ \bibinfo {pages} {057007} (\bibinfo {year}
		{2010})}\BibitemShut {NoStop}%
	\bibitem [{\citenamefont {Chen}\ \emph {et~al.}(2009)\citenamefont {Chen},
		\citenamefont {Hu}, \citenamefont {Luo},\ and\ \citenamefont
		{Wang}}]{chen09b}%
	\BibitemOpen
	\bibfield  {author} {\bibinfo {author} {\bibfnamefont {G.~F.}\ \bibnamefont
			{Chen}}, \bibinfo {author} {\bibfnamefont {W.~Z.}\ \bibnamefont {Hu}},
		\bibinfo {author} {\bibfnamefont {J.~L.}\ \bibnamefont {Luo}}, \ and\
		\bibinfo {author} {\bibfnamefont {N.~L.}\ \bibnamefont {Wang}},\ }\bibfield
	{title} {\enquote {\bibinfo {title} {{Multiple Phase Transitions in
					Single-Crystalline Na$_{1-\delta}$FeAs}},}\ }\href {\doibase
		10.1103/PhysRevLett.102.227004} {\bibfield  {journal} {\bibinfo  {journal}
			{Phys. Rev. Lett.}\ }\textbf {\bibinfo {volume} {102}},\ \bibinfo {pages}
		{227004} (\bibinfo {year} {2009})}\BibitemShut {NoStop}%
	\bibitem [{\citenamefont {Wang}\ \emph {et~al.}(2012)\citenamefont {Wang},
		\citenamefont {Luo}, \citenamefont {Yan}, \citenamefont {Ying}, \citenamefont
		{Xiang}, \citenamefont {Ye}, \citenamefont {Cheng}, \citenamefont {Li},
		\citenamefont {Hu},\ and\ \citenamefont {Chen}}]{wang12c}%
	\BibitemOpen
	\bibfield  {author} {\bibinfo {author} {\bibfnamefont {A.~F.}\ \bibnamefont
			{Wang}}, \bibinfo {author} {\bibfnamefont {X.~G.}\ \bibnamefont {Luo}},
		\bibinfo {author} {\bibfnamefont {Y.~J.}\ \bibnamefont {Yan}}, \bibinfo
		{author} {\bibfnamefont {J.~J.}\ \bibnamefont {Ying}}, \bibinfo {author}
		{\bibfnamefont {Z.~J.}\ \bibnamefont {Xiang}}, \bibinfo {author}
		{\bibfnamefont {G.~J.}\ \bibnamefont {Ye}}, \bibinfo {author} {\bibfnamefont
			{P.}~\bibnamefont {Cheng}}, \bibinfo {author} {\bibfnamefont {Z.~Y.}\
			\bibnamefont {Li}}, \bibinfo {author} {\bibfnamefont {W.~J.}\ \bibnamefont
			{Hu}}, \ and\ \bibinfo {author} {\bibfnamefont {X.~H.}\ \bibnamefont
			{Chen}},\ }\bibfield  {title} {\enquote {\bibinfo {title} {{Phase diagram and
					calorimetric properties of NaFe$_{1-x}$Co$_{x}$As}},}\ }\href {\doibase
		10.1103/PhysRevB.85.224521} {\bibfield  {journal} {\bibinfo  {journal} {Phys.
				Rev. B}\ }\textbf {\bibinfo {volume} {85}},\ \bibinfo {pages} {224521}
		(\bibinfo {year} {2012})}\BibitemShut {NoStop}%
	\bibitem [{\citenamefont {Wang}\ \emph {et~al.}(2013)\citenamefont {Wang},
		\citenamefont {Lin}, \citenamefont {Cheng}, \citenamefont {Ye}, \citenamefont
		{Chen}, \citenamefont {Ma}, \citenamefont {Lu}, \citenamefont {Lei},
		\citenamefont {Luo},\ and\ \citenamefont {Chen}}]{wang13}%
	\BibitemOpen
	\bibfield  {author} {\bibinfo {author} {\bibfnamefont {A.~F.}\ \bibnamefont
			{Wang}}, \bibinfo {author} {\bibfnamefont {J.~J.}\ \bibnamefont {Lin}},
		\bibinfo {author} {\bibfnamefont {P.}~\bibnamefont {Cheng}}, \bibinfo
		{author} {\bibfnamefont {G.~J.}\ \bibnamefont {Ye}}, \bibinfo {author}
		{\bibfnamefont {F.}~\bibnamefont {Chen}}, \bibinfo {author} {\bibfnamefont
			{J.~Q.}\ \bibnamefont {Ma}}, \bibinfo {author} {\bibfnamefont {X.~F.}\
			\bibnamefont {Lu}}, \bibinfo {author} {\bibfnamefont {B.}~\bibnamefont
			{Lei}}, \bibinfo {author} {\bibfnamefont {X.~G.}\ \bibnamefont {Luo}}, \ and\
		\bibinfo {author} {\bibfnamefont {X.~H.}\ \bibnamefont {Chen}},\ }\bibfield
	{title} {\enquote {\bibinfo {title} {{Phase diagram and physical properties
					of NaFe$_{1-x}$Cu$_x$As single crystals}},}\ }\href {\doibase
		10.1103/PhysRevB.88.094516} {\bibfield  {journal} {\bibinfo  {journal} {Phys.
				Rev. B}\ }\textbf {\bibinfo {volume} {88}},\ \bibinfo {pages} {094516}
		(\bibinfo {year} {2013})}\BibitemShut {NoStop}%
	\bibitem [{\citenamefont {Steckel}\ \emph {et~al.}(2015)\citenamefont
		{Steckel}, \citenamefont {Roslova}, \citenamefont {Beck}, \citenamefont
		{Morozov}, \citenamefont {Aswartham}, \citenamefont {Evtushinsky},
		\citenamefont {Blum}, \citenamefont {Abdel-Hafiez}, \citenamefont {Bombor},
		\citenamefont {Maletz}, \citenamefont {Borisenko}, \citenamefont {Shevelkov},
		\citenamefont {Wolter}, \citenamefont {Hess}, \citenamefont {Wurmehl},\ and\
		\citenamefont {B\"uchner}}]{steckel15}%
	\BibitemOpen
	\bibfield  {author} {\bibinfo {author} {\bibfnamefont {F.}~\bibnamefont
			{Steckel}}, \bibinfo {author} {\bibfnamefont {M.}~\bibnamefont {Roslova}},
		\bibinfo {author} {\bibfnamefont {R.}~\bibnamefont {Beck}}, \bibinfo {author}
		{\bibfnamefont {I.}~\bibnamefont {Morozov}}, \bibinfo {author} {\bibfnamefont
			{S.}~\bibnamefont {Aswartham}}, \bibinfo {author} {\bibfnamefont
			{D.}~\bibnamefont {Evtushinsky}}, \bibinfo {author} {\bibfnamefont
			{C.~G.~F.}\ \bibnamefont {Blum}}, \bibinfo {author} {\bibfnamefont
			{M.}~\bibnamefont {Abdel-Hafiez}}, \bibinfo {author} {\bibfnamefont
			{D.}~\bibnamefont {Bombor}}, \bibinfo {author} {\bibfnamefont
			{J.}~\bibnamefont {Maletz}}, \bibinfo {author} {\bibfnamefont
			{S.}~\bibnamefont {Borisenko}}, \bibinfo {author} {\bibfnamefont {A.~V.}\
			\bibnamefont {Shevelkov}}, \bibinfo {author} {\bibfnamefont {A.~U.~B.}\
			\bibnamefont {Wolter}}, \bibinfo {author} {\bibfnamefont {C.}~\bibnamefont
			{Hess}}, \bibinfo {author} {\bibfnamefont {S.}~\bibnamefont {Wurmehl}}, \
		and\ \bibinfo {author} {\bibfnamefont {B.}~\bibnamefont {B\"uchner}},\
	}\bibfield  {title} {\enquote {\bibinfo {title} {{Crystal growth and
					electronic phase diagram of 4d-doped Na$_{1-\delta}$Fe$_{1-x}$Rh$_x$As in
					comparison to 3d-doped Na$_{1-\delta}$Fe$_{1-x}$Co$_x$As}},}\ }\href
	{\doibase 10.1103/PhysRevB.91.184516} {\bibfield  {journal} {\bibinfo
			{journal} {Phys. Rev. B}\ }\textbf {\bibinfo {volume} {91}},\ \bibinfo
		{pages} {184516} (\bibinfo {year} {2015})}\BibitemShut {NoStop}%
	\bibitem [{\citenamefont {Kitagawa}\ \emph {et~al.}(2011)\citenamefont
		{Kitagawa}, \citenamefont {Mezaki}, \citenamefont {Matsubayashi},
		\citenamefont {Uwatoko},\ and\ \citenamefont {Takigawa}}]{kitagawa11}%
	\BibitemOpen
	\bibfield  {author} {\bibinfo {author} {\bibfnamefont {K.}~\bibnamefont
			{Kitagawa}}, \bibinfo {author} {\bibfnamefont {Y.}~\bibnamefont {Mezaki}},
		\bibinfo {author} {\bibfnamefont {K.}~\bibnamefont {Matsubayashi}}, \bibinfo
		{author} {\bibfnamefont {Y.}~\bibnamefont {Uwatoko}}, \ and\ \bibinfo
		{author} {\bibfnamefont {M.}~\bibnamefont {Takigawa}},\ }\bibfield  {title}
	{\enquote {\bibinfo {title} {{Crossover from Commensurate to Incommensurate
					Antiferromagnetism in Stoichiometric NaFeAs Revealed by Single-Crystal
					$^{23}$Na,$^{75}$As-NMR Experiments}},}\ }\href {\doibase
		10.1143/JPSJ.80.033705} {\bibfield  {journal} {\bibinfo  {journal} {J. Phys.
				Soc. Jpn.}\ }\textbf {\bibinfo {volume} {80}},\ \bibinfo {pages} {033705}
		(\bibinfo {year} {2011})}\BibitemShut {NoStop}%
	\bibitem [{\citenamefont {Ma}\ \emph {et~al.}(2011)\citenamefont {Ma},
		\citenamefont {Chen}, \citenamefont {Yao}, \citenamefont {Zhang},
		\citenamefont {Zhang}, \citenamefont {Xia},\ and\ \citenamefont
		{Yu}}]{ma11a}%
	\BibitemOpen
	\bibfield  {author} {\bibinfo {author} {\bibfnamefont {L.}~\bibnamefont
			{Ma}}, \bibinfo {author} {\bibfnamefont {G.~F.}\ \bibnamefont {Chen}},
		\bibinfo {author} {\bibfnamefont {D.-X.}\ \bibnamefont {Yao}}, \bibinfo
		{author} {\bibfnamefont {J.}~\bibnamefont {Zhang}}, \bibinfo {author}
		{\bibfnamefont {S.}~\bibnamefont {Zhang}}, \bibinfo {author} {\bibfnamefont
			{T.-L.}\ \bibnamefont {Xia}}, \ and\ \bibinfo {author} {\bibfnamefont
			{W.}~\bibnamefont {Yu}},\ }\bibfield  {title} {\enquote {\bibinfo {title}
			{{$^{23}$Na and $^{75}$As NMR study of antiferromagnetism and spin
					fluctuations in NaFeAs single crystals}},}\ }\href {\doibase
		10.1103/PhysRevB.83.132501} {\bibfield  {journal} {\bibinfo  {journal} {Phys.
				Rev. B}\ }\textbf {\bibinfo {volume} {83}},\ \bibinfo {pages} {132501}
		(\bibinfo {year} {2011})}\BibitemShut {NoStop}%
	\bibitem [{\citenamefont {Ning}\ \emph {et~al.}(2010)\citenamefont {Ning},
		\citenamefont {Ahilan}, \citenamefont {Imai}, \citenamefont {Sefat},
		\citenamefont {McGuire}, \citenamefont {Sales}, \citenamefont {Mandrus},
		\citenamefont {Cheng}, \citenamefont {Shen},\ and\ \citenamefont
		{Wen}}]{ning10}%
	\BibitemOpen
	\bibfield  {author} {\bibinfo {author} {\bibfnamefont {F.~L.}\ \bibnamefont
			{Ning}}, \bibinfo {author} {\bibfnamefont {K.}~\bibnamefont {Ahilan}},
		\bibinfo {author} {\bibfnamefont {T.}~\bibnamefont {Imai}}, \bibinfo {author}
		{\bibfnamefont {A.~S.}\ \bibnamefont {Sefat}}, \bibinfo {author}
		{\bibfnamefont {M.~A.}\ \bibnamefont {McGuire}}, \bibinfo {author}
		{\bibfnamefont {B.~C.}\ \bibnamefont {Sales}}, \bibinfo {author}
		{\bibfnamefont {D.}~\bibnamefont {Mandrus}}, \bibinfo {author} {\bibfnamefont
			{P.}~\bibnamefont {Cheng}}, \bibinfo {author} {\bibfnamefont
			{B.}~\bibnamefont {Shen}}, \ and\ \bibinfo {author} {\bibfnamefont {H.-H}\
			\bibnamefont {Wen}},\ }\bibfield  {title} {\enquote {\bibinfo {title}
			{{Contrasting Spin Dynamics between Underdoped and Overdoped
					Ba({Fe$_{1-x}$Co$_{x}$})$_2$As$_2$}},}\ }\href {\doibase
		10.1103/PhysRevLett.104.037001} {\bibfield  {journal} {\bibinfo  {journal}
			{Phys. Rev. Lett.}\ }\textbf {\bibinfo {volume} {104}},\ \bibinfo {pages}
		{037001} (\bibinfo {year} {2010})}\BibitemShut {NoStop}%
	\bibitem [{\citenamefont {Nakai}\ \emph {et~al.}(2013)\citenamefont {Nakai},
		\citenamefont {Iye}, \citenamefont {Kitagawa}, \citenamefont {Ishida},
		\citenamefont {Kasahara}, \citenamefont {Shibauchi}, \citenamefont {Matsuda},
		\citenamefont {Ikeda},\ and\ \citenamefont {Terashima}}]{nakai13}%
	\BibitemOpen
	\bibfield  {author} {\bibinfo {author} {\bibfnamefont {Y.}~\bibnamefont
			{Nakai}}, \bibinfo {author} {\bibfnamefont {T.}~\bibnamefont {Iye}}, \bibinfo
		{author} {\bibfnamefont {S.}~\bibnamefont {Kitagawa}}, \bibinfo {author}
		{\bibfnamefont {K.}~\bibnamefont {Ishida}}, \bibinfo {author} {\bibfnamefont
			{S.}~\bibnamefont {Kasahara}}, \bibinfo {author} {\bibfnamefont
			{T.}~\bibnamefont {Shibauchi}}, \bibinfo {author} {\bibfnamefont
			{Y.}~\bibnamefont {Matsuda}}, \bibinfo {author} {\bibfnamefont
			{H.}~\bibnamefont {Ikeda}}, \ and\ \bibinfo {author} {\bibfnamefont
			{T.}~\bibnamefont {Terashima}},\ }\bibfield  {title} {\enquote {\bibinfo
			{title} {{Normal-state spin dynamics in the iron-pnictide superconductors
					BaFe${}_{2}$(As${}_{1-x}$P${}_{x}$)${}_{2}$ and
					Ba(Fe${}_{1-x}$Co${}_{x}$)${}_{2}$As${}_{2}$ probed with NMR
					measurements}},}\ }\href {\doibase 10.1103/PhysRevB.87.174507} {\bibfield
		{journal} {\bibinfo  {journal} {Phys. Rev. B}\ }\textbf {\bibinfo {volume}
			{87}},\ \bibinfo {pages} {174507} (\bibinfo {year} {2013})}\BibitemShut
	{NoStop}%
	\bibitem [{\citenamefont {Takeda}\ \emph {et~al.}(2014)\citenamefont {Takeda},
		\citenamefont {Imai}, \citenamefont {Tachibana}, \citenamefont {Gaudet},
		\citenamefont {Gaulin}, \citenamefont {Saparov},\ and\ \citenamefont
		{Sefat}}]{takeda14}%
	\BibitemOpen
	\bibfield  {author} {\bibinfo {author} {\bibfnamefont {H.}~\bibnamefont
			{Takeda}}, \bibinfo {author} {\bibfnamefont {T.}~\bibnamefont {Imai}},
		\bibinfo {author} {\bibfnamefont {M.}~\bibnamefont {Tachibana}}, \bibinfo
		{author} {\bibfnamefont {J.}~\bibnamefont {Gaudet}}, \bibinfo {author}
		{\bibfnamefont {B.~D.}\ \bibnamefont {Gaulin}}, \bibinfo {author}
		{\bibfnamefont {B.~I.}\ \bibnamefont {Saparov}}, \ and\ \bibinfo {author}
		{\bibfnamefont {A.~S.}\ \bibnamefont {Sefat}},\ }\bibfield  {title} {\enquote
		{\bibinfo {title} {{Cu Substitution Effects on the Local Magnetic Properties
					of Ba(Fe$_{1-x}$Cu$_x$)$_2$As$_2$: A Site-Selective $^{75}$As and $^{63}$Cu
					NMR Study}},}\ }\href {\doibase 10.1103/PhysRevLett.113.117001} {\bibfield
		{journal} {\bibinfo  {journal} {Phys. Rev. Lett.}\ }\textbf {\bibinfo
			{volume} {113}},\ \bibinfo {pages} {117001} (\bibinfo {year}
		{2014})}\BibitemShut {NoStop}%
	\bibitem [{\citenamefont {Lifshitz}(1960)}]{lifshitz60}%
	\BibitemOpen
	\bibfield  {author} {\bibinfo {author} {\bibfnamefont {I.~M.}\ \bibnamefont
			{Lifshitz}},\ }\bibfield  {title} {\enquote {\bibinfo {title} {{Anomalies of
					electron characteristics of a metal in the high pressure region}},}\
	}\href@noop {} {\bibfield  {journal} {\bibinfo  {journal} {Sov. Phys. JETP}\
		}\textbf {\bibinfo {volume} {11}},\ \bibinfo {pages} {1130} (\bibinfo {year}
		{1960})}\BibitemShut {NoStop}%
	\bibitem [{\citenamefont {Nakai}\ \emph {et~al.}(2009)\citenamefont {Nakai},
		\citenamefont {Kitagawa}, \citenamefont {Ishida}, \citenamefont {Kamihara},
		\citenamefont {Hirano},\ and\ \citenamefont {Hosono}}]{nakai09}%
	\BibitemOpen
	\bibfield  {author} {\bibinfo {author} {\bibfnamefont {Y.}~\bibnamefont
			{Nakai}}, \bibinfo {author} {\bibfnamefont {S.}~\bibnamefont {Kitagawa}},
		\bibinfo {author} {\bibfnamefont {K.}~\bibnamefont {Ishida}}, \bibinfo
		{author} {\bibfnamefont {Y.}~\bibnamefont {Kamihara}}, \bibinfo {author}
		{\bibfnamefont {M.}~\bibnamefont {Hirano}}, \ and\ \bibinfo {author}
		{\bibfnamefont {H.}~\bibnamefont {Hosono}},\ }\bibfield  {title} {\enquote
		{\bibinfo {title} {{Systematic $^{75}$As NMR study of the dependence of
					low-lying excitations on F doping in the iron oxypnictide
					LaFeAsO$_{1-x}$F$_x$}},}\ }\href {\doibase 10.1103/PhysRevB.79.212506}
	{\bibfield  {journal} {\bibinfo  {journal} {Phys. Rev. B}\ }\textbf {\bibinfo
			{volume} {79}},\ \bibinfo {pages} {212506} (\bibinfo {year}
		{2009})}\BibitemShut {NoStop}%
	\bibitem [{\citenamefont {Ji}\ \emph {et~al.}(2013)\citenamefont {Ji},
		\citenamefont {Zhang}, \citenamefont {Ma}, \citenamefont {Fan}, \citenamefont
		{Wang}, \citenamefont {Dai}, \citenamefont {Tan}, \citenamefont {Song},
		\citenamefont {Zhang}, \citenamefont {Dai}, \citenamefont {Normand},\ and\
		\citenamefont {Yu}}]{ji13}%
	\BibitemOpen
	\bibfield  {author} {\bibinfo {author} {\bibfnamefont {G.~F.}\ \bibnamefont
			{Ji}}, \bibinfo {author} {\bibfnamefont {J.~S.}\ \bibnamefont {Zhang}},
		\bibinfo {author} {\bibfnamefont {Long}\ \bibnamefont {Ma}}, \bibinfo
		{author} {\bibfnamefont {P.}~\bibnamefont {Fan}}, \bibinfo {author}
		{\bibfnamefont {P.~S.}\ \bibnamefont {Wang}}, \bibinfo {author}
		{\bibfnamefont {J.}~\bibnamefont {Dai}}, \bibinfo {author} {\bibfnamefont
			{G.~T.}\ \bibnamefont {Tan}}, \bibinfo {author} {\bibfnamefont
			{Y.}~\bibnamefont {Song}}, \bibinfo {author} {\bibfnamefont {C.~L.}\
			\bibnamefont {Zhang}}, \bibinfo {author} {\bibfnamefont {P.}~\bibnamefont
			{Dai}}, \bibinfo {author} {\bibfnamefont {B.}~\bibnamefont {Normand}}, \ and\
		\bibinfo {author} {\bibfnamefont {W.}~\bibnamefont {Yu}},\ }\bibfield
	{title} {\enquote {\bibinfo {title} {{Simultaneous Optimization of Spin
					Fluctuations and Superconductivity under Pressure in an Iron-Based
					Superconductor}},}\ }\href {\doibase 10.1103/PhysRevLett.111.107004}
	{\bibfield  {journal} {\bibinfo  {journal} {Phys. Rev. Lett.}\ }\textbf
		{\bibinfo {volume} {111}},\ \bibinfo {pages} {107004} (\bibinfo {year}
		{2013})}\BibitemShut {NoStop}%
	\bibitem [{\citenamefont {Dioguardi}\ \emph {et~al.}(2013)\citenamefont
		{Dioguardi}, \citenamefont {Crocker}, \citenamefont {Shockley}, \citenamefont
		{Lin}, \citenamefont {Shirer}, \citenamefont {Nisson}, \citenamefont
		{Lawson}, \citenamefont {apRoberts Warren}, \citenamefont {Canfield},
		\citenamefont {Bud'ko}, \citenamefont {Ran},\ and\ \citenamefont
		{Curro}}]{dioguardi13}%
	\BibitemOpen
	\bibfield  {author} {\bibinfo {author} {\bibfnamefont {A.~P.}\ \bibnamefont
			{Dioguardi}}, \bibinfo {author} {\bibfnamefont {J.}~\bibnamefont {Crocker}},
		\bibinfo {author} {\bibfnamefont {A.~C.}\ \bibnamefont {Shockley}}, \bibinfo
		{author} {\bibfnamefont {C.~H.}\ \bibnamefont {Lin}}, \bibinfo {author}
		{\bibfnamefont {K.~R.}\ \bibnamefont {Shirer}}, \bibinfo {author}
		{\bibfnamefont {D.~M.}\ \bibnamefont {Nisson}}, \bibinfo {author}
		{\bibfnamefont {M.~M.}\ \bibnamefont {Lawson}}, \bibinfo {author}
		{\bibfnamefont {N.}~\bibnamefont {apRoberts Warren}}, \bibinfo {author}
		{\bibfnamefont {P.~C.}\ \bibnamefont {Canfield}}, \bibinfo {author}
		{\bibfnamefont {S.~L.}\ \bibnamefont {Bud'ko}}, \bibinfo {author}
		{\bibfnamefont {S.}~\bibnamefont {Ran}}, \ and\ \bibinfo {author}
		{\bibfnamefont {N.~J.}\ \bibnamefont {Curro}},\ }\bibfield  {title} {\enquote
		{\bibinfo {title} {{Coexistence of Cluster Spin Glass and Superconductivity
					in
					$\mathrm{Ba}({\mathrm{Fe}}_{1\mathbf{\ensuremath{-}}x}{\mathrm{Co}}_{x}{)}_{2}{\mathrm{As}}_{2}$
					for $0.060\mathbf{\ensuremath{\le}}x\mathbf{\ensuremath{\le}}0.071$}},}\
	}\href {\doibase 10.1103/PhysRevLett.111.207201} {\bibfield  {journal}
		{\bibinfo  {journal} {Phys. Rev. Lett.}\ }\textbf {\bibinfo {volume} {111}},\
		\bibinfo {pages} {207201} (\bibinfo {year} {2013})}\BibitemShut {NoStop}%
	\bibitem [{\citenamefont {Dioguardi}\ \emph {et~al.}(2015)\citenamefont
		{Dioguardi}, \citenamefont {Lawson}, \citenamefont {Bush}, \citenamefont
		{Crocker}, \citenamefont {Shirer}, \citenamefont {Nisson}, \citenamefont
		{Kissikov}, \citenamefont {Ran}, \citenamefont {Bud'ko}, \citenamefont
		{Canfield}, \citenamefont {Yuan}, \citenamefont {Kuhns}, \citenamefont
		{Reyes}, \citenamefont {Grafe},\ and\ \citenamefont {Curro}}]{dioguardi15}%
	\BibitemOpen
	\bibfield  {author} {\bibinfo {author} {\bibfnamefont {A.~P.}\ \bibnamefont
			{Dioguardi}}, \bibinfo {author} {\bibfnamefont {M.~M.}\ \bibnamefont
			{Lawson}}, \bibinfo {author} {\bibfnamefont {B.~T.}\ \bibnamefont {Bush}},
		\bibinfo {author} {\bibfnamefont {J.}~\bibnamefont {Crocker}}, \bibinfo
		{author} {\bibfnamefont {K.~R.}\ \bibnamefont {Shirer}}, \bibinfo {author}
		{\bibfnamefont {D.~M.}\ \bibnamefont {Nisson}}, \bibinfo {author}
		{\bibfnamefont {T.}~\bibnamefont {Kissikov}}, \bibinfo {author}
		{\bibfnamefont {S.}~\bibnamefont {Ran}}, \bibinfo {author} {\bibfnamefont
			{S.~L.}\ \bibnamefont {Bud'ko}}, \bibinfo {author} {\bibfnamefont {P.~C.}\
			\bibnamefont {Canfield}}, \bibinfo {author} {\bibfnamefont {S.}~\bibnamefont
			{Yuan}}, \bibinfo {author} {\bibfnamefont {P.~L.}\ \bibnamefont {Kuhns}},
		\bibinfo {author} {\bibfnamefont {A.~P.}\ \bibnamefont {Reyes}}, \bibinfo
		{author} {\bibfnamefont {H.-J.}\ \bibnamefont {Grafe}}, \ and\ \bibinfo
		{author} {\bibfnamefont {N.~J.}\ \bibnamefont {Curro}},\ }\bibfield  {title}
	{\enquote {\bibinfo {title} {{NMR evidence for inhomogeneous glassy behavior
					driven by nematic fluctuations in iron arsenide superconductors}},}\ }\href
	{\doibase 10.1103/PhysRevB.92.165116} {\bibfield  {journal} {\bibinfo
			{journal} {Phys. Rev. B}\ }\textbf {\bibinfo {volume} {92}},\ \bibinfo
		{pages} {165116} (\bibinfo {year} {2015})}\BibitemShut {NoStop}%
	\bibitem [{\citenamefont {Hunt}\ \emph {et~al.}(1999)\citenamefont {Hunt},
		\citenamefont {Singer}, \citenamefont {Thurber},\ and\ \citenamefont
		{Imai}}]{hunt99}%
	\BibitemOpen
	\bibfield  {author} {\bibinfo {author} {\bibfnamefont {A.~W.}\ \bibnamefont
			{Hunt}}, \bibinfo {author} {\bibfnamefont {P.~M.}\ \bibnamefont {Singer}},
		\bibinfo {author} {\bibfnamefont {K.~R.}\ \bibnamefont {Thurber}}, \ and\
		\bibinfo {author} {\bibfnamefont {T.}~\bibnamefont {Imai}},\ }\bibfield
	{title} {\enquote {\bibinfo {title} {{$^{63}$Cu NQR Measurement of Stripe
					Order Parameter in La$_{2-x}$Sr$_{x}$CuO$_4$}},}\ }\href {\doibase
		10.1103/PhysRevLett.82.4300} {\bibfield  {journal} {\bibinfo  {journal}
			{Phys. Rev. Lett.}\ }\textbf {\bibinfo {volume} {82}},\ \bibinfo {pages}
		{4300} (\bibinfo {year} {1999})}\BibitemShut {NoStop}%
	\bibitem [{\citenamefont {Imai}\ \emph {et~al.}(2017)\citenamefont {Imai},
		\citenamefont {Takahashi}, \citenamefont {Arsenault}, \citenamefont {Acton},
		\citenamefont {Lee}, \citenamefont {He}, \citenamefont {Lee},\ and\
		\citenamefont {Fujita}}]{imai17}%
	\BibitemOpen
	\bibfield  {author} {\bibinfo {author} {\bibfnamefont {T.}~\bibnamefont
			{Imai}}, \bibinfo {author} {\bibfnamefont {S.~K.}\ \bibnamefont {Takahashi}},
		\bibinfo {author} {\bibfnamefont {A.}~\bibnamefont {Arsenault}}, \bibinfo
		{author} {\bibfnamefont {A.~W.}\ \bibnamefont {Acton}}, \bibinfo {author}
		{\bibfnamefont {D.}~\bibnamefont {Lee}}, \bibinfo {author} {\bibfnamefont
			{W.}~\bibnamefont {He}}, \bibinfo {author} {\bibfnamefont {Y.~S.}\
			\bibnamefont {Lee}}, \ and\ \bibinfo {author} {\bibfnamefont
			{M.}~\bibnamefont {Fujita}},\ }\bibfield  {title} {\enquote {\bibinfo {title}
			{{Revisiting $^{63}\mathrm{Cu}$ NMR evidence for charge order in
					superconducting
					${\mathrm{La}}_{1.885}{\mathrm{Sr}}_{0.115}{\mathrm{CuO}}_{4}$}},}\ }\href
	{\doibase 10.1103/PhysRevB.96.224508} {\bibfield  {journal} {\bibinfo
			{journal} {Phys. Rev. B}\ }\textbf {\bibinfo {volume} {96}},\ \bibinfo
		{pages} {224508} (\bibinfo {year} {2017})}\BibitemShut {NoStop}%
	\bibitem [{\citenamefont {Imai}\ and\ \citenamefont {Lee}(2018)}]{imai18}%
	\BibitemOpen
	\bibfield  {author} {\bibinfo {author} {\bibfnamefont {T.}~\bibnamefont
			{Imai}}\ and\ \bibinfo {author} {\bibfnamefont {Y.~S.}\ \bibnamefont {Lee}},\
	}\bibfield  {title} {\enquote {\bibinfo {title} {{$^{139}\mathrm{La}$ and
					$^{63}\mathrm{Cu}$ NMR investigation of charge order in
					${\mathrm{La}}_{2}{\mathrm{CuO}}_{4+y}$ (${T}_{c}=42$ K)}},}\ }\href
	{\doibase 10.1103/PhysRevB.97.104506} {\bibfield  {journal} {\bibinfo
			{journal} {Phys. Rev. B}\ }\textbf {\bibinfo {volume} {97}},\ \bibinfo
		{pages} {104506} (\bibinfo {year} {2018})}\BibitemShut {NoStop}%
	\bibitem [{\citenamefont {Imai}\ \emph {et~al.}(2009)\citenamefont {Imai},
		\citenamefont {Ahilan}, \citenamefont {Ning}, \citenamefont {McQueen},\ and\
		\citenamefont {Cava}}]{imai09}%
	\BibitemOpen
	\bibfield  {author} {\bibinfo {author} {\bibfnamefont {T.}~\bibnamefont
			{Imai}}, \bibinfo {author} {\bibfnamefont {K.}~\bibnamefont {Ahilan}},
		\bibinfo {author} {\bibfnamefont {F.~L.}\ \bibnamefont {Ning}}, \bibinfo
		{author} {\bibfnamefont {T.~M.}\ \bibnamefont {McQueen}}, \ and\ \bibinfo
		{author} {\bibfnamefont {R.~J.}\ \bibnamefont {Cava}},\ }\bibfield  {title}
	{\enquote {\bibinfo {title} {{Why Does Undoped {FeSe} Become a High-{$T_c$}
					Superconductor under Pressure?}}}\ }\href {\doibase
		10.1103/PhysRevLett.102.177005} {\bibfield  {journal} {\bibinfo  {journal}
			{Phys. Rev. Lett.}\ }\textbf {\bibinfo {volume} {102}},\ \bibinfo {pages}
		{177005} (\bibinfo {year} {2009})}\BibitemShut {NoStop}%
	\bibitem [{\citenamefont {Wright}\ \emph {et~al.}(2012)\citenamefont {Wright},
		\citenamefont {Lancaster}, \citenamefont {Franke}, \citenamefont {Steele},
		\citenamefont {M\"oller}, \citenamefont {Pitcher}, \citenamefont {Corkett},
		\citenamefont {Parker}, \citenamefont {Free}, \citenamefont {Pratt},
		\citenamefont {Baker}, \citenamefont {Clarke},\ and\ \citenamefont
		{Blundell}}]{wright12}%
	\BibitemOpen
	\bibfield  {author} {\bibinfo {author} {\bibfnamefont {J.~D.}\ \bibnamefont
			{Wright}}, \bibinfo {author} {\bibfnamefont {T.}~\bibnamefont {Lancaster}},
		\bibinfo {author} {\bibfnamefont {I.}~\bibnamefont {Franke}}, \bibinfo
		{author} {\bibfnamefont {A.~J.}\ \bibnamefont {Steele}}, \bibinfo {author}
		{\bibfnamefont {J.~S.}\ \bibnamefont {M\"oller}}, \bibinfo {author}
		{\bibfnamefont {M.~J.}\ \bibnamefont {Pitcher}}, \bibinfo {author}
		{\bibfnamefont {A.~J.}\ \bibnamefont {Corkett}}, \bibinfo {author}
		{\bibfnamefont {D.~R.}\ \bibnamefont {Parker}}, \bibinfo {author}
		{\bibfnamefont {D.~G.}\ \bibnamefont {Free}}, \bibinfo {author}
		{\bibfnamefont {F.~L.}\ \bibnamefont {Pratt}}, \bibinfo {author}
		{\bibfnamefont {P.~J.}\ \bibnamefont {Baker}}, \bibinfo {author}
		{\bibfnamefont {S.~J.}\ \bibnamefont {Clarke}}, \ and\ \bibinfo {author}
		{\bibfnamefont {S.~J.}\ \bibnamefont {Blundell}},\ }\bibfield  {title}
	{\enquote {\bibinfo {title} {{Gradual destruction of magnetism in the
					superconducting family NaFe$_{1-x}$Co$_x$As}},}\ }\href {\doibase
		10.1103/PhysRevB.85.054503} {\bibfield  {journal} {\bibinfo  {journal} {Phys.
				Rev. B}\ }\textbf {\bibinfo {volume} {85}},\ \bibinfo {pages} {054503}
		(\bibinfo {year} {2012})}\BibitemShut {NoStop}%
	\bibitem [{\citenamefont {Zhou}\ \emph {et~al.}(2016)\citenamefont {Zhou},
		\citenamefont {Xing}, \citenamefont {Wang}, \citenamefont {Jin},\ and\
		\citenamefont {Zheng}}]{zhou16}%
	\BibitemOpen
	\bibfield  {author} {\bibinfo {author} {\bibfnamefont {R.}~\bibnamefont
			{Zhou}}, \bibinfo {author} {\bibfnamefont {L.~Y.}\ \bibnamefont {Xing}},
		\bibinfo {author} {\bibfnamefont {X.~C.}\ \bibnamefont {Wang}}, \bibinfo
		{author} {\bibfnamefont {C.~Q.}\ \bibnamefont {Jin}}, \ and\ \bibinfo
		{author} {\bibfnamefont {Guo-qing}\ \bibnamefont {Zheng}},\ }\bibfield
	{title} {\enquote {\bibinfo {title} {{Orbital order and spin nematicity in
					the tetragonal phase of the electron-doped iron pnictides
					NaFe$_{1-x}$Co$_x$As}},}\ }\href {\doibase 10.1103/PhysRevB.93.060502}
	{\bibfield  {journal} {\bibinfo  {journal} {Phys. Rev. B}\ }\textbf {\bibinfo
			{volume} {93}},\ \bibinfo {pages} {060502} (\bibinfo {year}
		{2016})}\BibitemShut {NoStop}%
	\bibitem [{\citenamefont {Luetkens}\ \emph {et~al.}(2009)\citenamefont
		{Luetkens}, \citenamefont {Klauss}, \citenamefont {Kraken}, \citenamefont
		{Litterst}, \citenamefont {Dellmann}, \citenamefont {Klingeler},
		\citenamefont {Hess}, \citenamefont {Khasanov}, \citenamefont {Amato},
		\citenamefont {Baines}, \citenamefont {Kosmala}, \citenamefont {Schumann},
		\citenamefont {Braden}, \citenamefont {Hamann-Borrero}, \citenamefont {Leps},
		\citenamefont {Kondrat}, \citenamefont {Behr}, \citenamefont {Werner},\ and\
		\citenamefont {B\"uchner}}]{luetkens09}%
	\BibitemOpen
	\bibfield  {author} {\bibinfo {author} {\bibfnamefont {H.}~\bibnamefont
			{Luetkens}}, \bibinfo {author} {\bibfnamefont {H.-H.}\ \bibnamefont
			{Klauss}}, \bibinfo {author} {\bibfnamefont {M.}~\bibnamefont {Kraken}},
		\bibinfo {author} {\bibfnamefont {F.~J.}\ \bibnamefont {Litterst}}, \bibinfo
		{author} {\bibfnamefont {T.}~\bibnamefont {Dellmann}}, \bibinfo {author}
		{\bibfnamefont {R.}~\bibnamefont {Klingeler}}, \bibinfo {author}
		{\bibfnamefont {C.}~\bibnamefont {Hess}}, \bibinfo {author} {\bibfnamefont
			{R.}~\bibnamefont {Khasanov}}, \bibinfo {author} {\bibfnamefont
			{A.}~\bibnamefont {Amato}}, \bibinfo {author} {\bibfnamefont
			{C.}~\bibnamefont {Baines}}, \bibinfo {author} {\bibfnamefont
			{M.}~\bibnamefont {Kosmala}}, \bibinfo {author} {\bibfnamefont {O.~J.}\
			\bibnamefont {Schumann}}, \bibinfo {author} {\bibfnamefont {M.}~\bibnamefont
			{Braden}}, \bibinfo {author} {\bibfnamefont {J.}~\bibnamefont
			{Hamann-Borrero}}, \bibinfo {author} {\bibfnamefont {N.}~\bibnamefont
			{Leps}}, \bibinfo {author} {\bibfnamefont {A.}~\bibnamefont {Kondrat}},
		\bibinfo {author} {\bibfnamefont {G.}~\bibnamefont {Behr}}, \bibinfo {author}
		{\bibfnamefont {J.}~\bibnamefont {Werner}}, \ and\ \bibinfo {author}
		{\bibfnamefont {B.}~\bibnamefont {B\"uchner}},\ }\bibfield  {title} {\enquote
		{\bibinfo {title} {{The electronic phase diagram of the LaO$_{1-x}$F$_x$FeAs
					superconductor}},}\ }\href {\doibase 10.1038/nmat2397} {\bibfield  {journal}
		{\bibinfo  {journal} {Nature Mater.}\ }\textbf {\bibinfo {volume} {8}},\
		\bibinfo {pages} {305} (\bibinfo {year} {2009})}\BibitemShut {NoStop}%
	\bibitem [{\citenamefont {Ma}\ \emph {et~al.}(2014)\citenamefont {Ma},
		\citenamefont {Dai}, \citenamefont {Wang}, \citenamefont {Lu}, \citenamefont
		{Song}, \citenamefont {Zhang}, \citenamefont {Tan}, \citenamefont {Dai},
		\citenamefont {Hu}, \citenamefont {Li}, \citenamefont {Normand},\ and\
		\citenamefont {Yu}}]{ma14}%
	\BibitemOpen
	\bibfield  {author} {\bibinfo {author} {\bibfnamefont {L.}~\bibnamefont
			{Ma}}, \bibinfo {author} {\bibfnamefont {J.}~\bibnamefont {Dai}}, \bibinfo
		{author} {\bibfnamefont {P.~S.}\ \bibnamefont {Wang}}, \bibinfo {author}
		{\bibfnamefont {X.~R.}\ \bibnamefont {Lu}}, \bibinfo {author} {\bibfnamefont
			{Y.}~\bibnamefont {Song}}, \bibinfo {author} {\bibfnamefont {C.}~\bibnamefont
			{Zhang}}, \bibinfo {author} {\bibfnamefont {G.~T.}\ \bibnamefont {Tan}},
		\bibinfo {author} {\bibfnamefont {P.}~\bibnamefont {Dai}}, \bibinfo {author}
		{\bibfnamefont {D.}~\bibnamefont {Hu}}, \bibinfo {author} {\bibfnamefont
			{S.~L.}\ \bibnamefont {Li}}, \bibinfo {author} {\bibfnamefont
			{B.}~\bibnamefont {Normand}}, \ and\ \bibinfo {author} {\bibfnamefont
			{W.}~\bibnamefont {Yu}},\ }\bibfield  {title} {\enquote {\bibinfo {title}
			{{Phase separation, competition, and volume-fraction control in
					NaFe$_{1-x}$Co$_x$As}},}\ }\href {\doibase 10.1103/PhysRevB.90.144502}
	{\bibfield  {journal} {\bibinfo  {journal} {Phys. Rev. B}\ }\textbf {\bibinfo
			{volume} {90}},\ \bibinfo {pages} {144502} (\bibinfo {year}
		{2014})}\BibitemShut {NoStop}%
	\bibitem [{\citenamefont {Fernandes}\ \emph {et~al.}(2012)\citenamefont
		{Fernandes}, \citenamefont {Chubukov}, \citenamefont {Knolle}, \citenamefont
		{Eremin},\ and\ \citenamefont {Schmalian}}]{fernandes12}%
	\BibitemOpen
	\bibfield  {author} {\bibinfo {author} {\bibfnamefont {R.~M.}\ \bibnamefont
			{Fernandes}}, \bibinfo {author} {\bibfnamefont {A.~V.}\ \bibnamefont
			{Chubukov}}, \bibinfo {author} {\bibfnamefont {J.}~\bibnamefont {Knolle}},
		\bibinfo {author} {\bibfnamefont {I.}~\bibnamefont {Eremin}}, \ and\ \bibinfo
		{author} {\bibfnamefont {J.}~\bibnamefont {Schmalian}},\ }\bibfield  {title}
	{\enquote {\bibinfo {title} {{Preemptive nematic order, pseudogap, and
					orbital order in the iron pnictides}},}\ }\href {\doibase
		10.1103/PhysRevB.85.024534} {\bibfield  {journal} {\bibinfo  {journal} {Phys.
				Rev. B}\ }\textbf {\bibinfo {volume} {85}},\ \bibinfo {pages} {024534}
		(\bibinfo {year} {2012})}\BibitemShut {NoStop}%
	\bibitem [{\citenamefont {Chubukov}\ \emph {et~al.}(2008)\citenamefont
		{Chubukov}, \citenamefont {Efremov},\ and\ \citenamefont
		{Eremin}}]{chubukov08}%
	\BibitemOpen
	\bibfield  {author} {\bibinfo {author} {\bibfnamefont {A.~V.}\ \bibnamefont
			{Chubukov}}, \bibinfo {author} {\bibfnamefont {D.~V.}\ \bibnamefont
			{Efremov}}, \ and\ \bibinfo {author} {\bibfnamefont {I.}~\bibnamefont
			{Eremin}},\ }\bibfield  {title} {\enquote {\bibinfo {title} {Magnetism,
				superconductivity, and pairing symmetry in iron-based superconductors},}\
	}\href {\doibase 10.1103/PhysRevB.78.134512} {\bibfield  {journal} {\bibinfo
			{journal} {Phys. Rev. B}\ }\textbf {\bibinfo {volume} {78}},\ \bibinfo
		{pages} {134512} (\bibinfo {year} {2008})}\BibitemShut {NoStop}%
	\bibitem [{\citenamefont {Kang}\ and\ \citenamefont
		{Te\v{s}anovi\'{c}}(2011)}]{kang11}%
	\BibitemOpen
	\bibfield  {author} {\bibinfo {author} {\bibfnamefont {J.}~\bibnamefont
			{Kang}}\ and\ \bibinfo {author} {\bibfnamefont {Z.}~\bibnamefont
			{Te\v{s}anovi\'{c}}},\ }\bibfield  {title} {\enquote {\bibinfo {title}
			{Theory of the valley-density wave and hidden order in iron pnictides},}\
	}\href {\doibase 10.1103/PhysRevB.83.020505} {\bibfield  {journal} {\bibinfo
			{journal} {Phys. Rev. B}\ }\textbf {\bibinfo {volume} {83}},\ \bibinfo
		{pages} {020505} (\bibinfo {year} {2011})}\BibitemShut {NoStop}%
	\bibitem [{\citenamefont {Classen}\ \emph {et~al.}(2017)\citenamefont
		{Classen}, \citenamefont {Xing}, \citenamefont {Khodas},\ and\ \citenamefont
		{Chubukov}}]{classen17}%
	\BibitemOpen
	\bibfield  {author} {\bibinfo {author} {\bibfnamefont {L.}~\bibnamefont
			{Classen}}, \bibinfo {author} {\bibfnamefont {R.-Q.}\ \bibnamefont {Xing}},
		\bibinfo {author} {\bibfnamefont {M.}~\bibnamefont {Khodas}}, \ and\ \bibinfo
		{author} {\bibfnamefont {A.~V.}\ \bibnamefont {Chubukov}},\ }\bibfield
	{title} {\enquote {\bibinfo {title} {{Interplay between Magnetism,
					Superconductivity, and Orbital Order in 5-Pocket Model for Iron-Based
					Superconductors: Parquet Renormalization Group Study}},}\ }\href {\doibase
		10.1103/PhysRevLett.118.037001} {\bibfield  {journal} {\bibinfo  {journal}
			{Phys. Rev. Lett.}\ }\textbf {\bibinfo {volume} {118}},\ \bibinfo {pages}
		{037001} (\bibinfo {year} {2017})}\BibitemShut {NoStop}%
\end{thebibliography}

\clearpage

\begin{thebibliography}{2}%
\makeatletter
\providecommand \@ifxundefined [1]{%
 \@ifx{#1\undefined}
}%
\providecommand \@ifnum [1]{%
 \ifnum #1\expandafter \@firstoftwo
 \else \expandafter \@secondoftwo
 \fi
}%
\providecommand \@ifx [1]{%
 \ifx #1\expandafter \@firstoftwo
 \else \expandafter \@secondoftwo
 \fi
}%
\providecommand \natexlab [1]{#1}%
\providecommand \enquote  [1]{``#1''}%
\providecommand \bibnamefont  [1]{#1}%
\providecommand \bibfnamefont [1]{#1}%
\providecommand \citenamefont [1]{#1}%
\providecommand \href@noop [0]{\@secondoftwo}%
\providecommand \href [0]{\begingroup \@sanitize@url \@href}%
\providecommand \@href[1]{\@@startlink{#1}\@@href}%
\providecommand \@@href[1]{\endgroup#1\@@endlink}%
\providecommand \@sanitize@url [0]{\catcode `\\12\catcode `\$12\catcode
  `\&12\catcode `\#12\catcode `\^12\catcode `\_12\catcode `\%12\relax}%
\providecommand \@@startlink[1]{}%
\providecommand \@@endlink[0]{}%
\providecommand \url  [0]{\begingroup\@sanitize@url \@url }%
\providecommand \@url [1]{\endgroup\@href {#1}{\urlprefix }}%
\providecommand \urlprefix  [0]{URL }%
\providecommand \Eprint [0]{\href }%
\providecommand \doibase [0]{http://dx.doi.org/}%
\providecommand \selectlanguage [0]{\@gobble}%
\providecommand \bibinfo  [0]{\@secondoftwo}%
\providecommand \bibfield  [0]{\@secondoftwo}%
\providecommand \translation [1]{[#1]}%
\providecommand \BibitemOpen [0]{}%
\providecommand \bibitemStop [0]{}%
\providecommand \bibitemNoStop [0]{.\EOS\space}%
\providecommand \EOS [0]{\spacefactor3000\relax}%
\providecommand \BibitemShut  [1]{\csname bibitem#1\endcsname}%
\let\auto@bib@innerbib\@empty
\bibitem [{\citenamefont {Parker}\ \emph {et~al.}(2009)\citenamefont {Parker},
  \citenamefont {Pitcher}, \citenamefont {Baker}, \citenamefont {Franke},
  \citenamefont {Lancaster}, \citenamefont {Blundell},\ and\ \citenamefont
  {Clarke}}]{parker09}%
  \BibitemOpen
  \bibfield  {author} {\bibinfo {author} {\bibfnamefont {D.~R.}\ \bibnamefont
  {Parker}}, \bibinfo {author} {\bibfnamefont {M.~J.}\ \bibnamefont {Pitcher}},
  \bibinfo {author} {\bibfnamefont {P.~J.}\ \bibnamefont {Baker}}, \bibinfo
  {author} {\bibfnamefont {I.}~\bibnamefont {Franke}}, \bibinfo {author}
  {\bibfnamefont {T.}~\bibnamefont {Lancaster}}, \bibinfo {author}
  {\bibfnamefont {S.~J.}\ \bibnamefont {Blundell}}, \ and\ \bibinfo {author}
  {\bibfnamefont {S.~J.}\ \bibnamefont {Clarke}},\ }\bibfield  {title}
  {\enquote {\bibinfo {title} {{Structure, antiferromagnetism and
  superconductivity of the layered iron arsenide NaFeAs}},}\ }\href {\doibase
  10.1039/B818911K} {\bibfield  {journal} {\bibinfo  {journal} {Chem. Commun.}\
  ,\ \bibinfo {pages} {2189--2191}} (\bibinfo {year} {2009})}\BibitemShut
  {NoStop}%
\bibitem [{\citenamefont {Ning}\ \emph {et~al.}(2010)\citenamefont {Ning},
  \citenamefont {Ahilan}, \citenamefont {Imai}, \citenamefont {Sefat},
  \citenamefont {McGuire}, \citenamefont {Sales}, \citenamefont {Mandrus},
  \citenamefont {Cheng}, \citenamefont {Shen},\ and\ \citenamefont
  {Wen}}]{ning10}%
  \BibitemOpen
  \bibfield  {author} {\bibinfo {author} {\bibfnamefont {F.~L.}\ \bibnamefont
  {Ning}}, \bibinfo {author} {\bibfnamefont {K.}~\bibnamefont {Ahilan}},
  \bibinfo {author} {\bibfnamefont {T.}~\bibnamefont {Imai}}, \bibinfo {author}
  {\bibfnamefont {A.~S.}\ \bibnamefont {Sefat}}, \bibinfo {author}
  {\bibfnamefont {M.~A.}\ \bibnamefont {McGuire}}, \bibinfo {author}
  {\bibfnamefont {B.~C.}\ \bibnamefont {Sales}}, \bibinfo {author}
  {\bibfnamefont {D.}~\bibnamefont {Mandrus}}, \bibinfo {author} {\bibfnamefont
  {P.}~\bibnamefont {Cheng}}, \bibinfo {author} {\bibfnamefont
  {B.}~\bibnamefont {Shen}}, \ and\ \bibinfo {author} {\bibfnamefont {H.-H}\
  \bibnamefont {Wen}},\ }\bibfield  {title} {\enquote {\bibinfo {title}
  {{Contrasting Spin Dynamics between Underdoped and Overdoped
  Ba({Fe$_{1-x}$Co$_{x}$})$_2$As$_2$}},}\ }\href {\doibase
  10.1103/PhysRevLett.104.037001} {\bibfield  {journal} {\bibinfo  {journal}
  {Phys. Rev. Lett.}\ }\textbf {\bibinfo {volume} {104}},\ \bibinfo {pages}
  {037001} (\bibinfo {year} {2010})}\BibitemShut {NoStop}%
\end{thebibliography}

%


\pagebreak

\paragraph*{\bf Acknowledgments}  This work was financially supported by National Creative Research Initiative (2010-0018300) and Global Collaborative Research Projects (2016K1A4A3914691) through Korea's NRF, which is funded by Ministry of Science, ICT and Future Planning (MSIP).
    The work at Germany has been supported by the Deutsche
		Forschungsgemeinschaft (Germany) via DFG Research Grants BA 4927/2-1. DVE acknowledges VW-foundation for partial financial support.
\paragraph*{\bf Author Contributions}  KHK and BSL have proposed and initiated the
	project.  WHN, BSL, DB have grown single crystals and characterized
		transport and structure properties. SHB performed NMR
		measurements and analyzed data;
    SHB, DVE, and KHK participated in writing of the manuscript. All authors
		discussed the results and commented on the manuscript.
\paragraph*{\bf Competing financial interests}  The Authors declare no Competing Financial or Non-Financial Interests.
\paragraph*{\bf Additional information} Correspondence and requests for materials should be
	addressed to S.-H. Baek~(email: sbaek.fu@gmail.com) or K. H.
		Kim~(email: optopia@snu.ac.kr).

\clearpage

\begin{figure}
	\centering
\includegraphics[width=\linewidth]{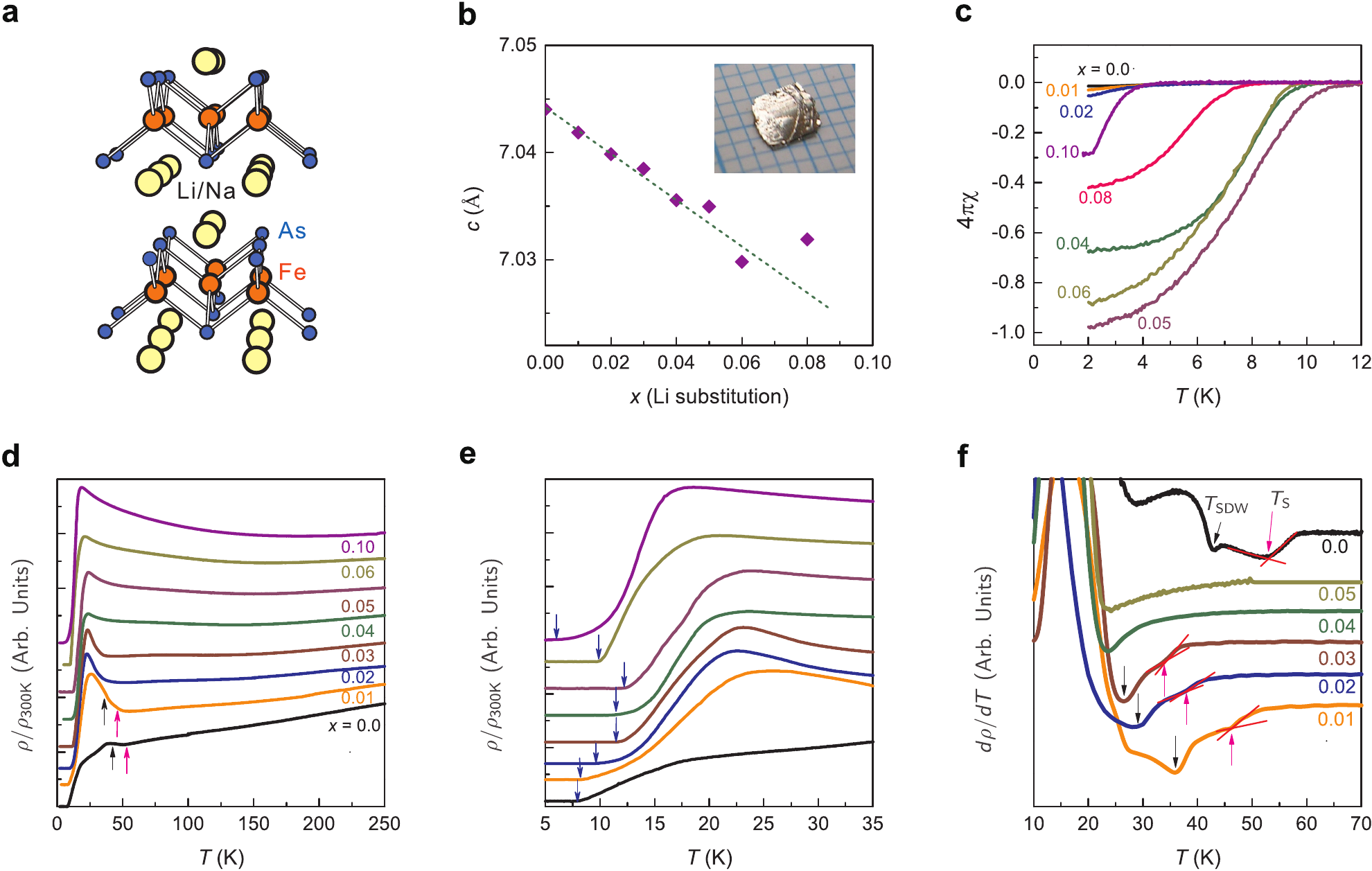}
\caption{Characterization of the \nlfa\ single crystals. 
	\bc{a}, Crystal structure of \nlfa.
	\bc{b}, Variation of the $c$-axis lattice parameter of the Na$_{1-x}$Li$_{x}$FeAs
		single crystals at 300 K  with a linear dashed guide line. Inset shows the $ab$-plane image of a single crystal in a millimeter scale.
	\bc{c}, Zero-field cooled magnetizations at low
	temperatures at $H_{ab}$ = 10 Oe, revealing SC transition and shielding fractions.  \bc{d-f},
	Temperature dependence of normalized resistivity, $\rho/\rho_\text{300K}$,
	enlarged resistivity near the SC
	transition, and $d\rho/dT$ for selected samples. Black and magenta
	arrows in \bc{d} and \bc{f} denote SDW ($T_\text{SDW}$) and structural ($T_S$)
	transitions, respectively, while blue arrows in \bc{e} reflect the $T_\text{c}^\rho$
	evolution with $x$.
	 Note that although previous studies of
		NaFeAs have observed a local inflection point in the $\rho$ curve at \tsdw\ (or a sharp dip in $d\rho/dT$),
		we observed weak minimum for our NaFeAs due to the presence of twinning.}
\label{fig1}
\end{figure}

\begin{figure}
\centering
\includegraphics[width=\linewidth]{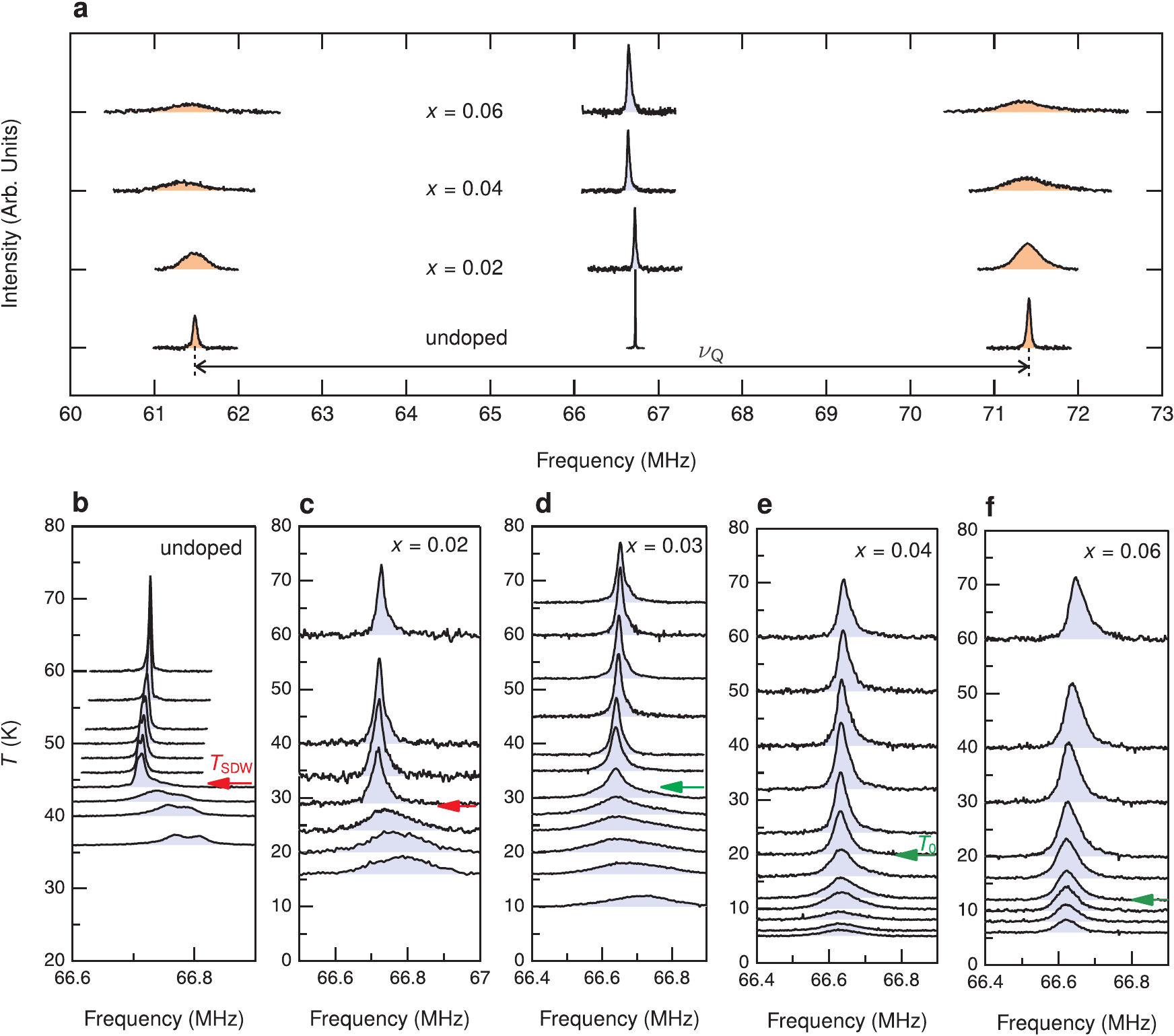}
\caption{\as\ NMR spectra in \nlfa\ for for $H \parallel ab$.
	\bc{a}, Central and satellite lines of \as\ as a function of doping at
	a fixed temperature of 60 K at 9.1 T. While the linewidth of the
	spectra increases with increasing doping, there is no other lines found and the
	quadrupole frequency $\nu_\text{Q}$ remains nearly unchanged.
	\bc{b}-\bc{f}, Temperature dependence of the \as\ central line as a
	function of $x$. For $x=0$ and 0.02, the \as\ spectrum starts to
	broaden and shift to higher frequency below \tsdw. In
	contrast, for $x\geq 0.04$, neither a significant broadening nor a shift of the line
	was observed, evidencing the absence of static magnetic
	moments. An inhomogeneous line broadening below $T_0$ for the intermediate doping $x=0.03$ is ascribed to  short-range magnetism associated with the charge/orbital ordered phase.
	The large reduction of the signal intensity below $\sim 10$ K is due to
	bulk superconductivity.
}
\label{spec_ab}
\end{figure}

\pagebreak

\begin{figure}
\centering
\includegraphics[width=\linewidth]{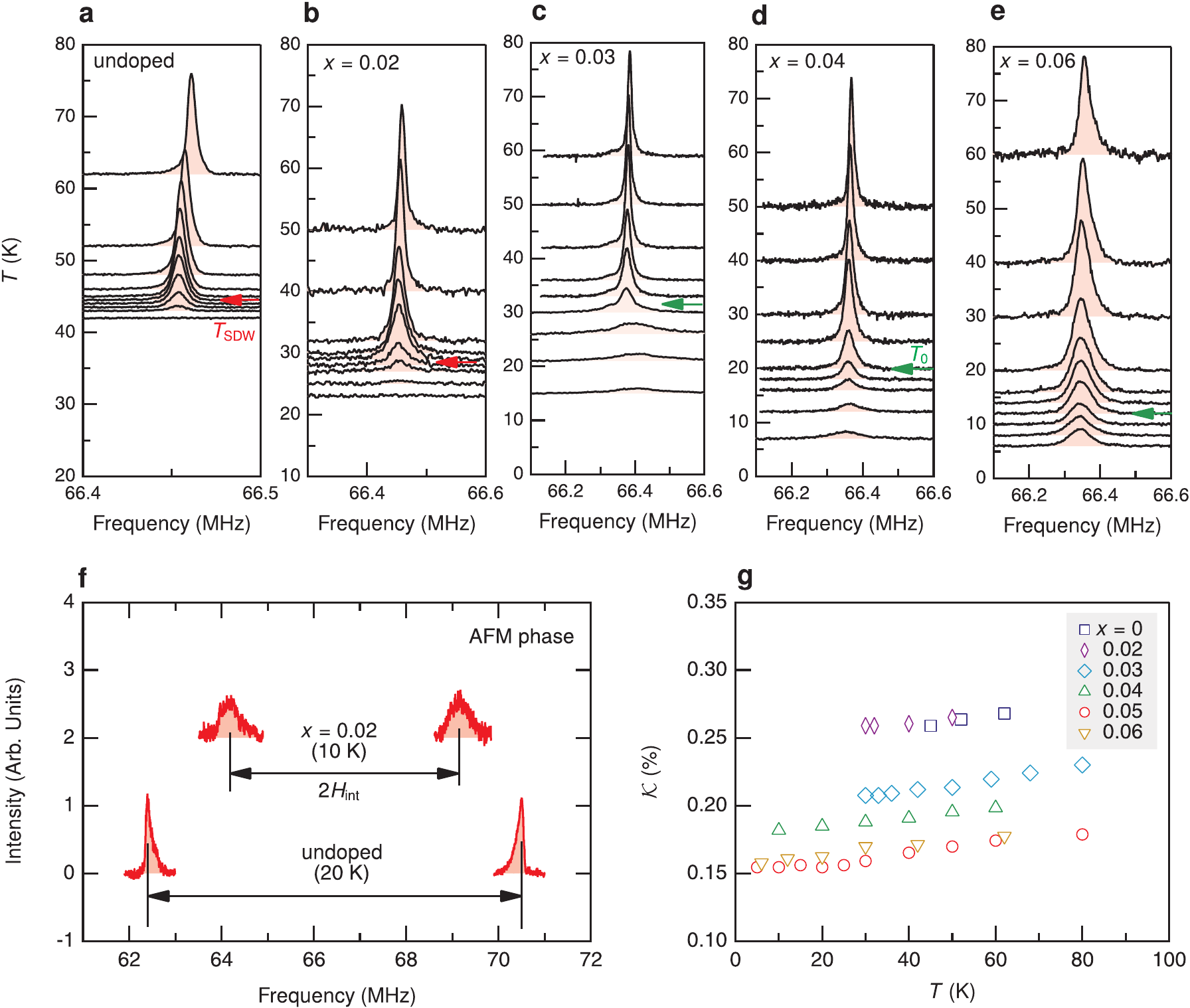}
\caption{\as\ NMR spectra in \nlfa\ for $H\parallel c$. 
	\bc{a}-\bc{e}, Temperature dependence of the central line as a
	function of Li doping $x$ measured at 9.1 T. For $x=0$ and 0.02, the \as\ line rapidly
	disappears below \tsdw\ due to the large hyperfine field resulting from the
	static Fe spin moment arranged antiferromagentically.
	For $x\geq 0.03$, the \as\ intensity remains
	finite down to low temperatures without a significant broadening, indicating the absence of long range magnetic order.
	\bc{f}, The AFM split \as\ lines were detected at low
	temperatures for $x=0$ and 0.02, manifesting the commensurate AFM phase.
	\bc{g}, The Knight shift $\mathcal{K}$
	as a function of temperature
	and doping at 9.1 T parallel to $c$.  The almost constant $\mathcal{K}$
	for a given $x$ is rapidly reduced for $x\geq 0.03$,
	suggesting a change of the Fermi surface geometry.
	}
\label{spec_c}
\end{figure}

\pagebreak

\begin{figure}
\centering
\includegraphics[width=0.7\linewidth]{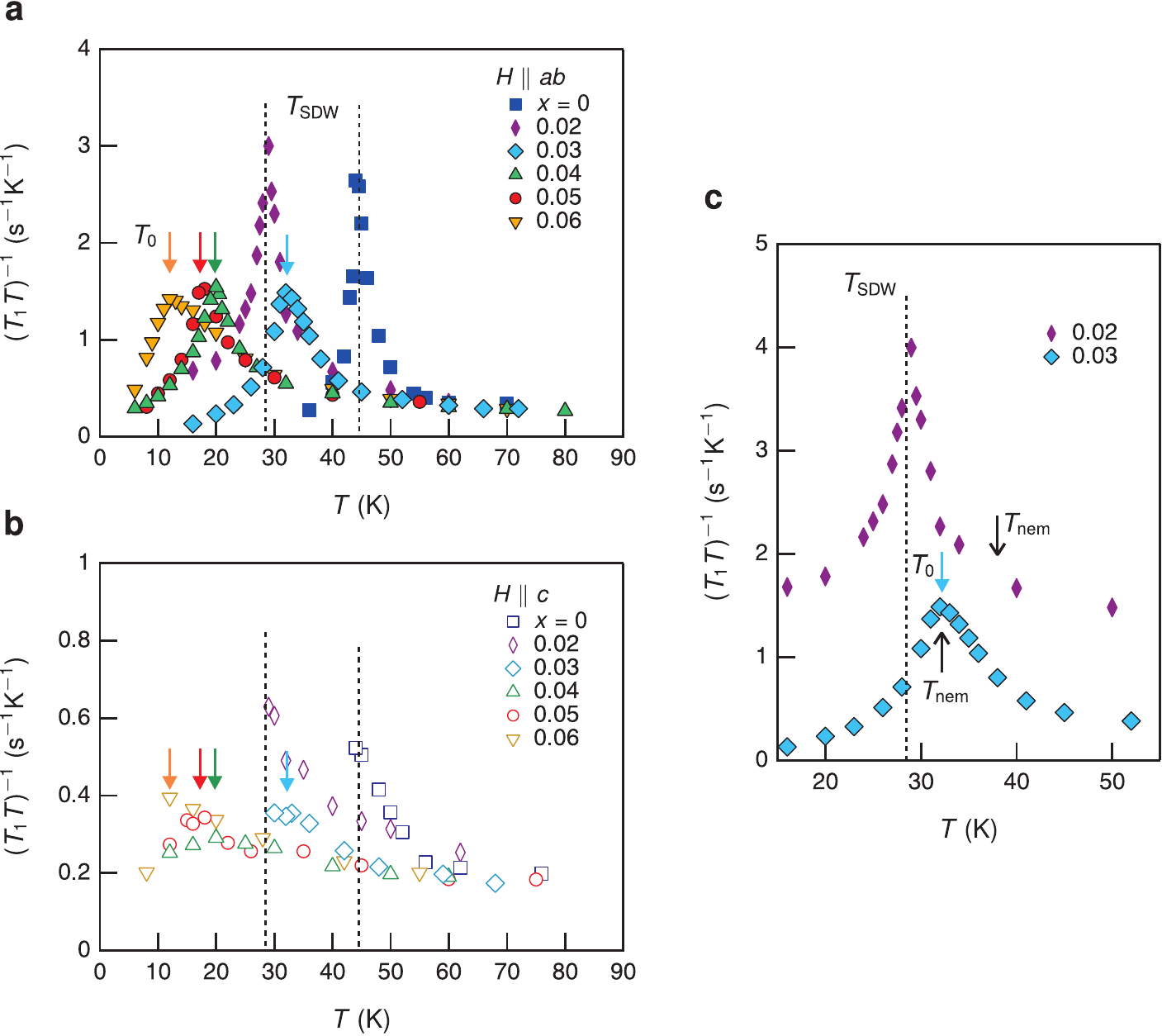}
\caption{Spin-lattice relaxation rates as a function
	of temperature and doping.
	\bc{a} and \bc{b}, The spin-lattice relaxation rate divided by
	temperature, \slrt, measured at 9.1 T perpendicular and parallel
	to $c$, respectively. For $x=0$ and 0.02, the SDW transitions were
	identified by the sharp peak of \slrt\ at \tsdw. Similar sharp
	transitions were observed for larger dopings ($x\geq 0.03$) at
	temperatures denoted by $T_0$, without an indication of
	long range SDW order.
	The comparison between \bc{a} and \bc{b}
	reveals that the strong anisotropy of spin fluctuations persists up to $x=0.06$.
	\bc{c}, Comparison between the \slrt\ data  for
	$x=0.02$ and $x=0.03$ obtained with $H\parallel ab$. The data for $x=0.02$
	are offset vertically for clarity. It shows that $T_0$
	for $x=0.03$ is higher than \tsdw\ for $x=0.02$, but nearly
	coincides with \tnem\ obtained by resistivity (see Fig.
	\ref{fig1}).
}
\label{t1t}
\end{figure}

\begin{figure}
\centering
\includegraphics[width=0.6\linewidth]{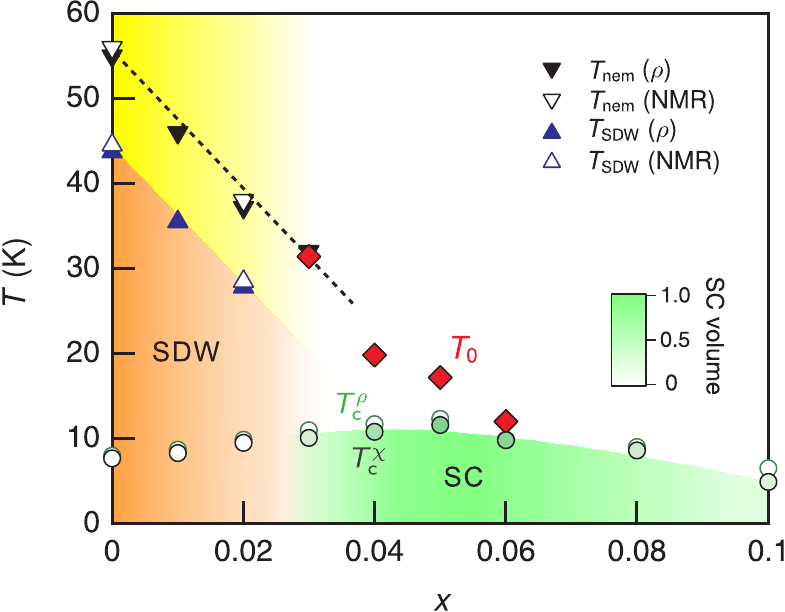}
\caption{Temperature-doping phase diagram of \nlfa.
\tsdw, \tnem, $T_\text{c}^\rho$, and $T_\text{c}^\chi$ were obtained by our
transport, magnetization, and NMR measurements which turn out to be consistent one
another. The emerging phase at $T_0$ before entering the bulk SC state is seemingly connected to the nematic transition at lower dopings. The shade of green below $T_\text{c}$ schematically represents the SC volume
	fraction obtained from Fig. \ref{fig1}c.
}
\label{phase}
\end{figure}

\clearpage

\renewcommand{\figurename}{{\bf Supplementary Figure}}
\renewcommand{\thefigure}{{\bf \arabic{figure}}}
\setcounter{figure}{0}

\begin{figure}[h]
	\centering
	\includegraphics[width=.75\linewidth]{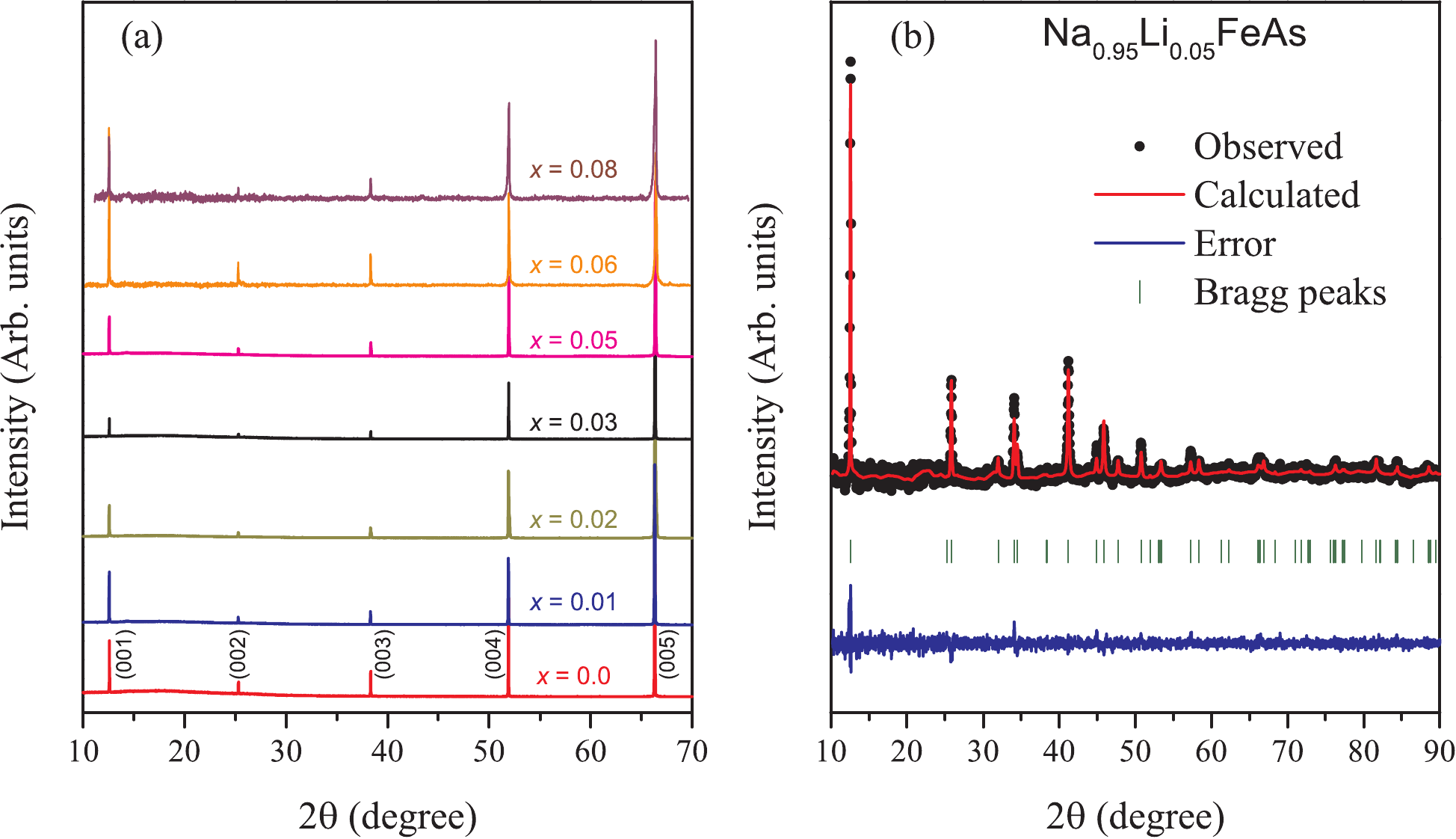}
	\caption{X-ray diffraction data. (a) The single crystalline x-ray diffraction data of selected  Na$_{1-x}$Li$_x$FeAs single crystals. (00$l$) reflection peaks can only be seen in the XRD patterns suggesting the absence of any other impurity phase. The diffraction patterns could be successfully refined by the tetragonal $P$4/$nmm$ structure as in the parent NaFeAs (ref. \onlinecite{parker09}) (b) Powder diffraction pattern of a ground Na$_{0.95}$Li$_{0.05}$FeAs single crystals.}
	\label{xrd}
\end{figure}

\begin{figure}[h]
	\centering
	\includegraphics[width=.75\linewidth]{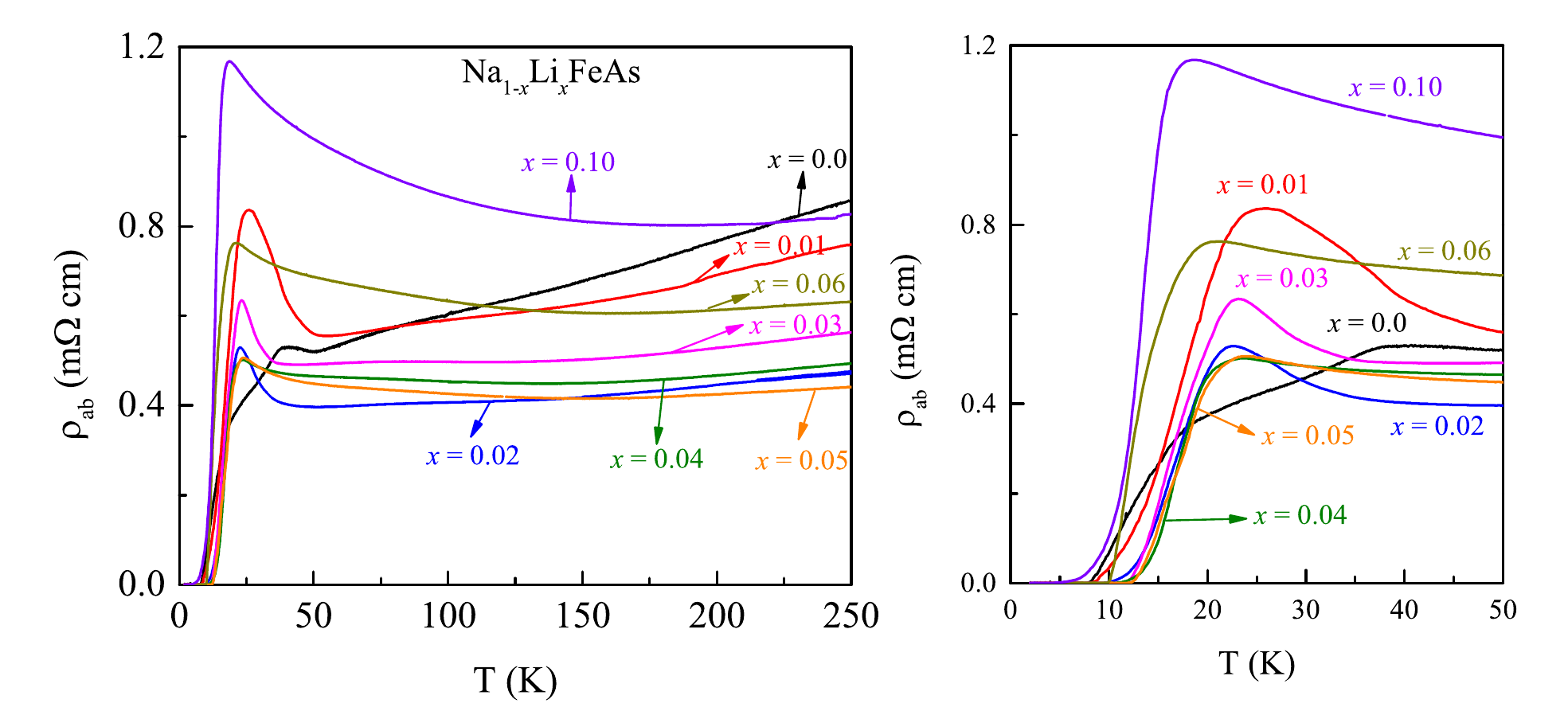}
	\caption{Original resistivity data. Temperature dependence of the in-plane resistivity in Na$_{1-x}$Li$_x$FeAs single crystals for different Li concentration $x$.}
	\label{Res}
\end{figure}

\begin{figure*}[h]
	\centering
	\includegraphics[width=.65\linewidth]{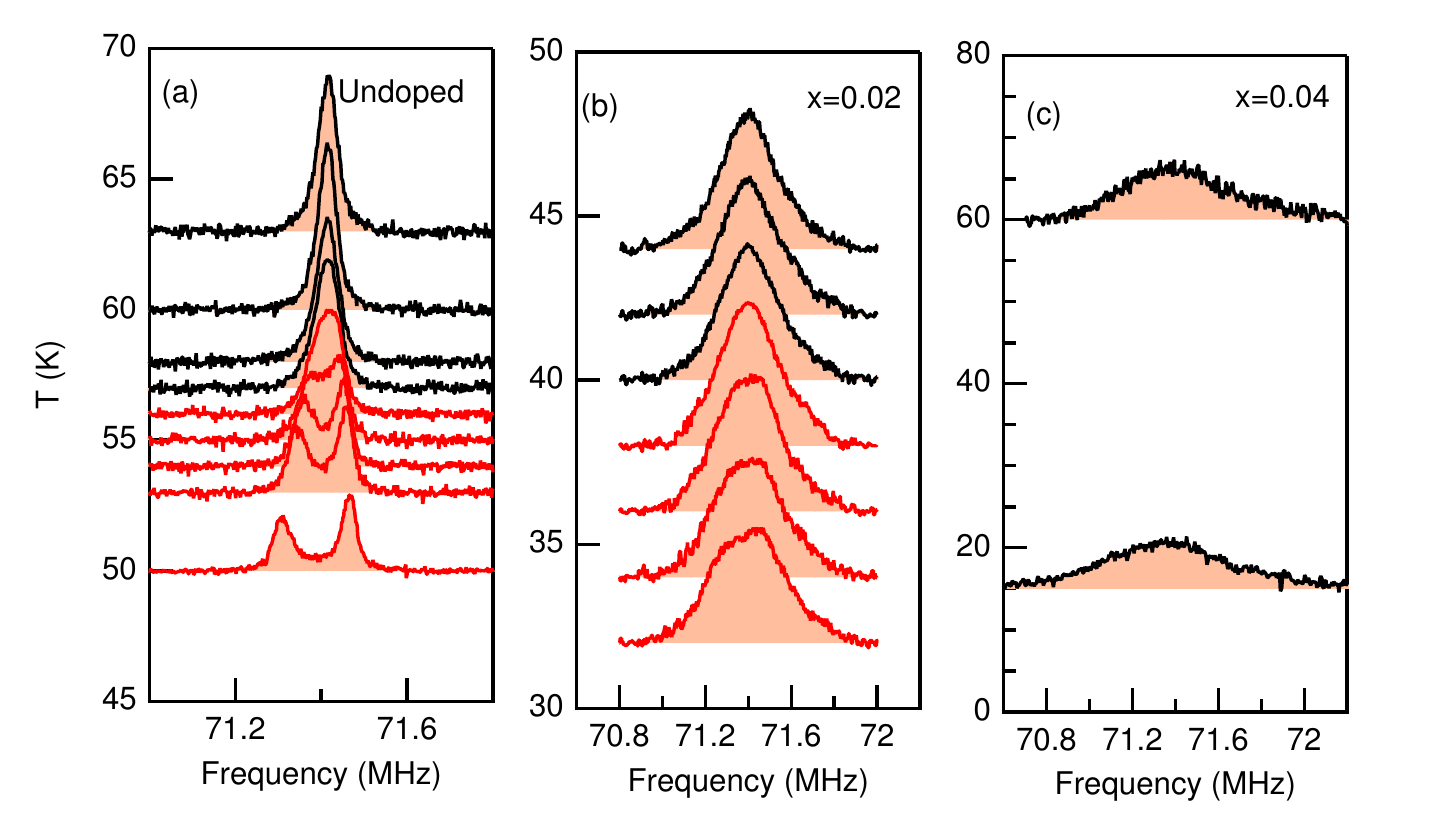}
	\caption{Determination of the nematic transition temperature by NMR. Temperature dependence of the $^{75}$As satellite line for (a) $x=0$, (b) 0.02, and (c) 0.04. The line splitting caused by nematic order was observed for $x=0$ and 0.02, which allows us to determine the nematic transition temperature. For $x=0.04$, the splitting is not detected, probably due to the larger broadening of the line than the splitting.}
	\label{S3}
\end{figure*}

\begin{figure*}[h]
	\centering
	\includegraphics[width=.75\linewidth]{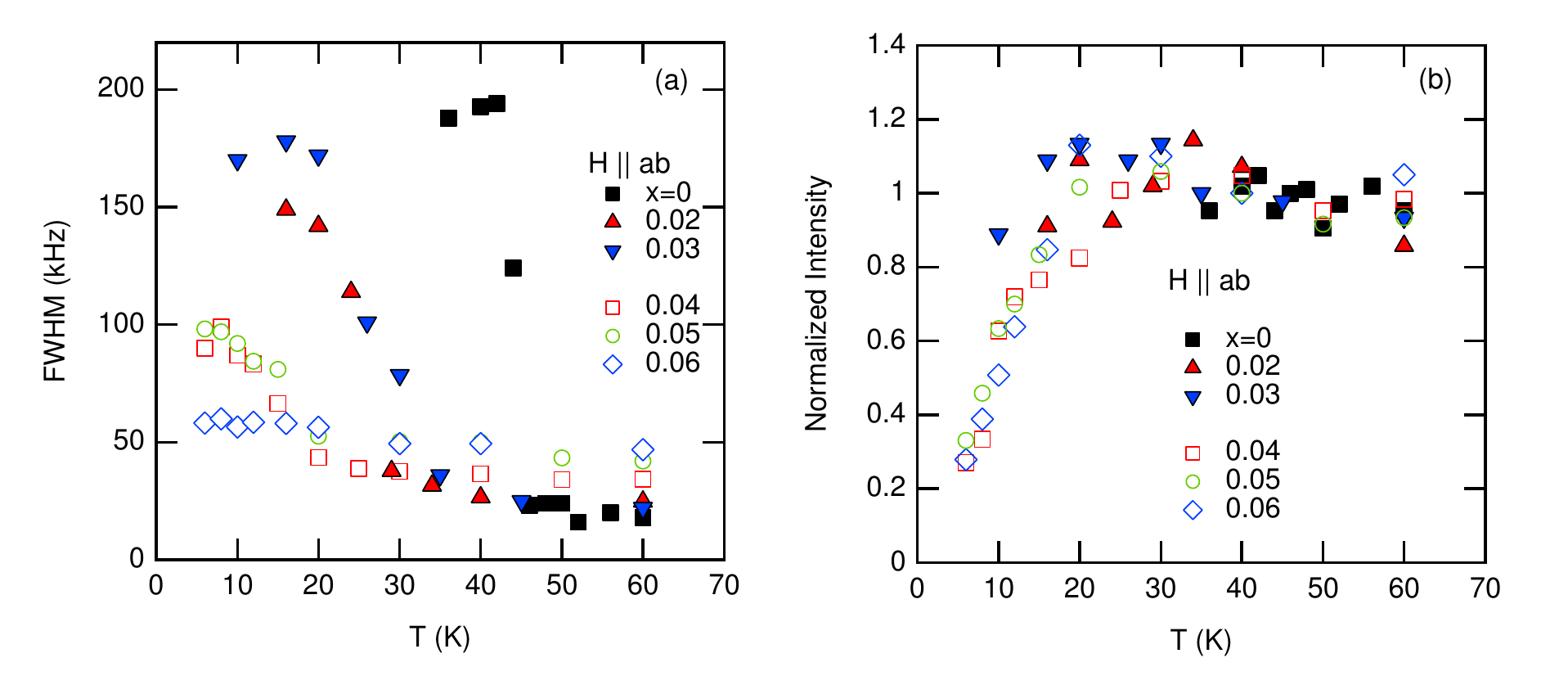}
	\caption{Detailed analysis of $^{75}$As spectra. (a) The full width at half maximum (FWHM) as a function of doping and temperature. For $x\geq 0.04$, the FWHM is weakly broadened down to low temperatures. Note that for $x=0.06$ there is no broadening of the line at all, while the sharp peak of $(T_1T)^{-1}$ is clearly observed.  For $x=0$, the FWHM data were multiplied by 4 to compare directly with those of doped samples. (b) Normalized signal intensity as a function of doping and temperature. While the signal intensity  does not change within experimental error for $x< 0.03$, it is notably suppressed below $T_0$ for $x\geq 0.04$.  }
	\label{S1}
\end{figure*}

\newpage

\begin{figure*}[h]
	\centering
	\includegraphics[width=.55\linewidth]{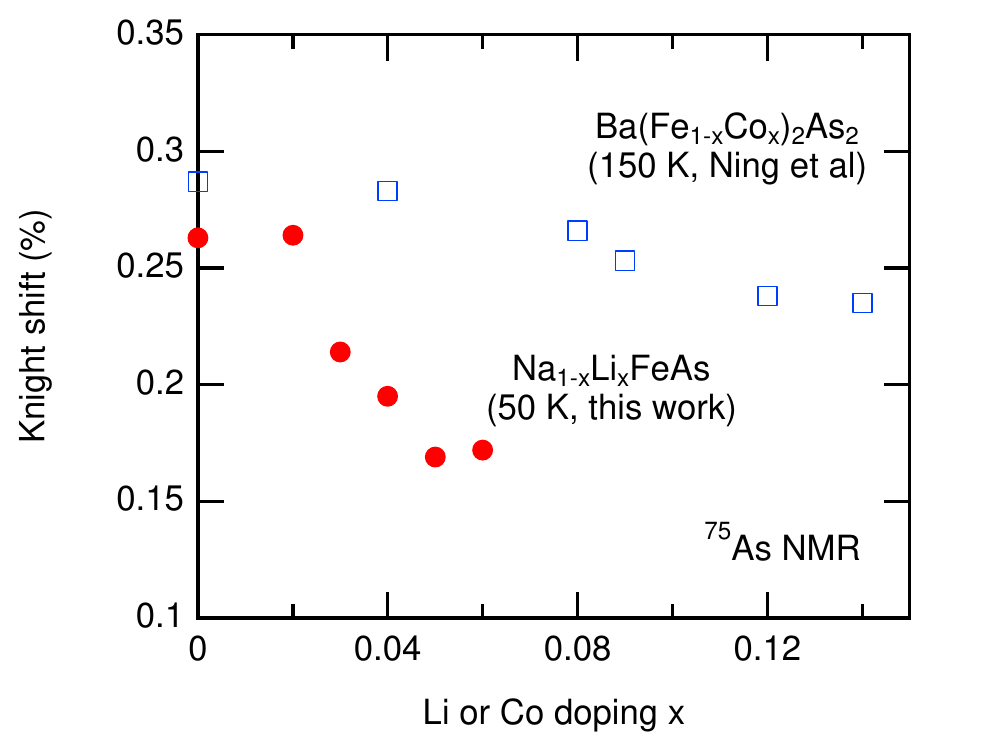}
	\caption{Sharp reduction of the Knight shift near $x=0.03$. Doping dependence of the Knight shift $\mathcal{K}$ at 50 K is compared with data in Ba(Fe$_{1-x}$Co$_x$)$_2$As$_2$ at 150 K (taken from ref. \onlinecite{ning10}).}
	\label{S2}
\end{figure*}

\begin{figure*}[h]
	\centering
	\includegraphics[width=.65\linewidth]{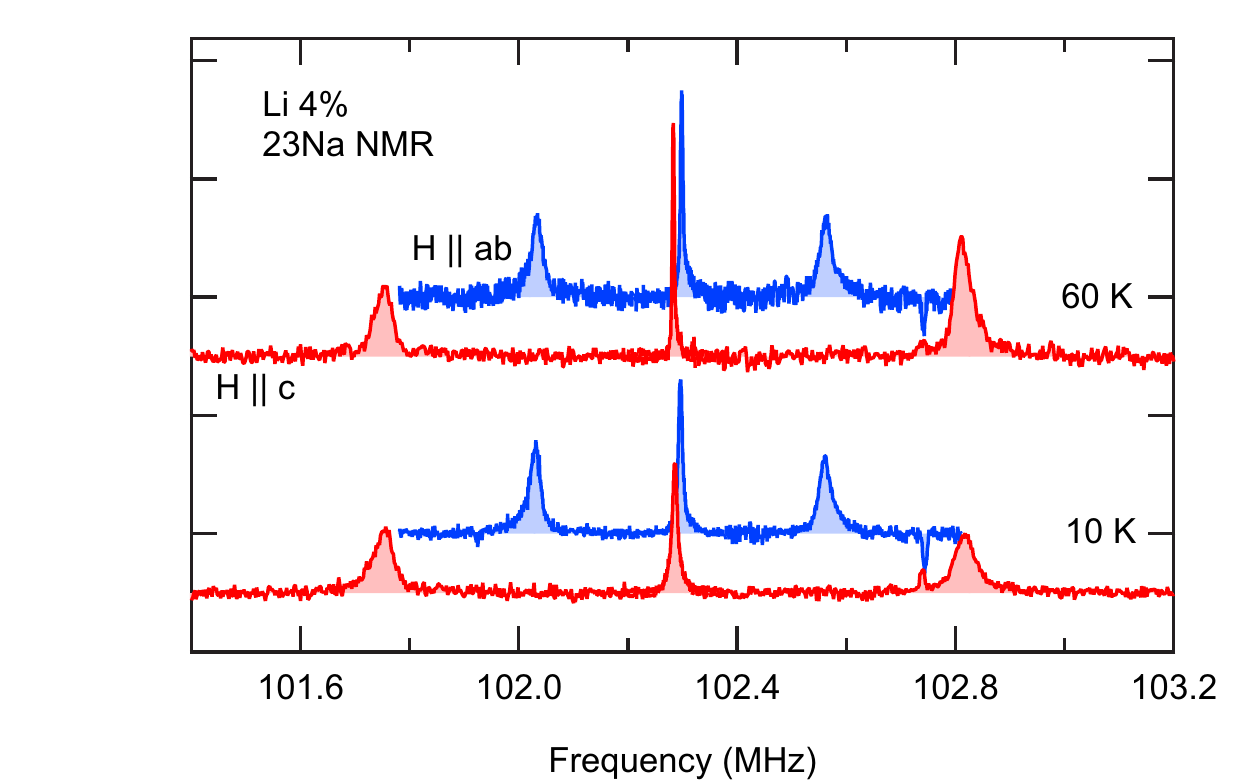}
	\caption{Comparison of $^{23}$Na spectra above and below $T_0$ for $x=0.04$. The spectrum does not reveal a notable difference in the ordered phase, unlike the $^{75}$As results, reflecting that the $^{75}$As effectively probes the Fe sites due to the much stronger hyperfine coupling.}
	\label{S4}
\end{figure*}

\end{document}